\documentclass[preprint,5p,twocolumn]{elsarticle}

\usepackage{amssymb}
\usepackage{amsmath}
\usepackage{xspace}
\usepackage{subfig}
\usepackage{color}
\usepackage{graphicx}
\usepackage{multirow}
\usepackage{lineno}
\usepackage{units}
\usepackage{hyperref}
\usepackage{cleveref}

\crefname{figure}{Fig.}{Figs.}
\crefname{equation}{Eq.}{Eqs.}
\crefname{section}{Section}{Sections}
\crefname{appendix}{Appendix}{Appendices}
\def \xmax {$X_{\rm max}$\xspace}
\def \xmaxmean {$\langle X_{\rm max} \rangle$\xspace}
\def \xmaxrms {$\sigma[X_{\rm max}]$\xspace}

\def \nmuexp{$N_{\mu}^{\rm meas}$\xspace}
\def \nmutotal{$N_{\mu}^{\rm tot}$\xspace}

\def \lgnmuexp{$\log_{10}N_{\mu}^{\rm meas}$\xspace}

\def \nmu {$N_{\mu}$\xspace}
\def \lgnmu {$ \log _{10}N_{\mu}$\xspace}

\def \lgnmumean {$\langle \log _{10}N_{\mu} \rangle$\xspace}
\def \lgnmurms {$\sigma[\log _{10}N_{\mu}]$\xspace}

\def \lgnmumeanexp {$\langle \log _{10}N_{\mu}^{\rm meas} \rangle$\xspace}
\def \lgnmurmsexp {$\sigma[\log _{10}N_{\mu}^{\rm meas}]$\xspace}

\def \lge{$\log_{10}(E)$\xspace}
\def \lna{$\ln(A)$\xspace}

\def \cchi{$\chi^2$\xspace}

\def \chiminalpha{$\chi^2_{\rm min}(a)$\xspace}
\def \chiminbeta{$\chi^2_{\rm min}(b)$\xspace}

\def \fnmu{$\alpha_{\rm N_{\mu}}$\xspace}
\def \fen{$\alpha_{\rm E}$\xspace}

\def\eV{\ifmmode {\mathrm{\ e\kern -0.1em V}}\else\textrm{e\kern -0.1em V}\fi\xspace}%
\newcommand{\expo}[1]{\unit[$^{#1}$]}
\newcommand{\energyEV}[1]{\unit[$10^{#1}$]{\eV}}
\newcommand{\energy}[1]{$10^{#1}$}
\journal{Astroparticle Physics}

\begin{document}

%\begin{linenumbers}

\begin{frontmatter}

\title{Interpretation of measurements of the number of muons in extensive air shower experiments}

\author[1]{Raul R. Prado}
\author[2]{Ruben Concei\c{c}\~ao}
\author[2]{M\'ario Pimenta}
\author[1]{Vitor de Souza}
\address[1]{Instituto de F\'isica de S\~ao Carlos, Universidade de S\~ao Paulo, S\~ao Carlos, Brazil}
\address[2]{Laborat\'orio de Instrumenta\c{c}\~ao e F\'isica Experimental de Part\'iculas, Lisbon, Portugal}

\begin{abstract}

  In this paper we analyze the energy evolution of the muon content of air showers between $10^{18.4}$ and $10^{19.6}$ eV to be able to determine the most likely mass composition scenario from future number of muons measurements. The energy and primary mass evolution of the number of muons is studied based on the Heitler-Matthews model and Monte Carlo simulation of the air shower. A simple model to describe the evolution of the first and second moments of number of muons distributions is proposed and validated. An analysis approach based on the comparison between this model's predictions and data to discriminate among a set of composition scenarios is presented and tested with simulations. It is shown that the composition scenarios can be potentially discriminated under the conditions imposed by the method. The discrimination power of the proposed analysis is stable under systematic changes of the absolute number of muons from model predictions and on the scale of the reconstructed energy.

\end{abstract}


\end{frontmatter}

%\newpage
%\tableofcontents

\newpage
%================================================================
\section{Introduction}
\label{sec:intro}

%%% UHECR astrophysics

The energy spectrum of ultra high energy cosmic rays (UHECRs) has been measured recently with high precision and two major features were confirmed. The ankle ($\log (E/{\rm eV}) \sim 18.7$) and the flux suppression ($\log (E/{\rm eV}) \sim 19.5$) have been undoubtedly established by HiRes~\cite{Abbasi2010a}, the Pierre Auger Observatory~\cite{Abraham2010e,AugerSpec2013} and Telescope Array~\cite{Abu-Zayyad2013a}. However, the astrophysical interpretation of these structures cannot be inferred with complete certainty mainly because of the lack of knowledge on the UHECR composition at these energies. In a light abundance scenario, the ankle could be interpreted as the modulation resulting from the particle interaction with radiation backgrounds \cite{Allard2005a,Berezinsky2006}. On the other hand, it could also be explained as the transition from galactic to extra-galactic cosmic rays \cite{Allard2015}. The flux suppression can be equally well described by the energy losses of extra-galactic particles due to interactions with CMB photons \cite{Berezinsky2005} or by the maximum reachable energy of the astrophysical acceleration mechanisms in nearby sources \cite{Biermann2012}. In each one of these astrophysical scenarios, the energy evolution of the UHECR composition is significantly different.

% The knowledge of the mass abundance of UHECRs is also important for understanding the acceleration and propagation mechanisms generating astroparticles.

%%% measurements and composition

The UHECR measurements are done indirectly through the detection of extensive air showers. Therefore, the determination of the composition depends strongly on the data analysis capability to correlate the measured properties of the shower to the primary particle type. This correlation is achieved using air shower simulations. However, intrinsic fluctuations of the showers and uncertainties in the high energy hadronic interaction models for energies above \energyEV{17} prevent us from a definitive conclusion about the primary particle type for each event. Statistical analysis and evolution trends~\cite{Abraham2010b,Aab2014} are used to minimize the fluctuation effects, nevertheless an unique interpretation of the data is not possible because of the hadronic interaction model uncertainties. Currently, the most reliable observable to investigate composition at higher energies is \xmax, the atmospheric depth at which the shower reaches the maximum number of particles~\cite{Kampert2012}. A second very powerful observable sensitive to primary particle mass is the number of muons (\nmu) in the showers. However, the lack of knowledge of the high energy hadronic interactions and the systematic uncertainties in the energy determination limit the interpretation of \nmu data in terms of composition in a more severe way than they do for \xmax. There are several indications that the current most often used hadronic interaction models fail at predicting the muonic component features of air showers~\cite{Farrar2013,Aab2015a}. Moreover, as \nmu scales directly with shower energy, the systematic uncertainty in energy reconstruction (typically $\sim 10 - 20\%$) represents also a difficult challenge to overcome in the interpretation of the \nmu data. As a consequence, it is not straightforward to envisage a data analysis procedure that extracts the mass abundance from the \nmu data.

%Despite the very accurate measurements of the \xmax distribution for energies above \energyEV{17.8} released by the Pierre Auger Collaboration~\cite{Aab2014}, the composition interpretation of the data depends on the hadronic interaction model~\cite{Aab2014a}.

%%% composition and Nmu

% similarly to what is done for \xmax~\cite{Aab2014a}.

%%% nao sei onde deixar esse paragrafo (??)
%Hadronic interaction properties and mass abundance are usually interrelated problems in the study of UHECRs. Supposing the mass abundance were known \textit{a priori}, one could make inferences about the hadronic interaction properties from the air shower measurements. On the other hand, if the uncertainties on the hadronic interaction models were smaller, the interpretation of the data in terms of mass abundance would be more precise.

%%%%%%% method and paper structure

In this paper we propose a new approach to interpret \nmu data which accommodates the systematic uncertainties of the high energy hadronic interaction models and of the energy reconstruction. The analysis proposed here is based on the energy evolution of the first (\lgnmumean) and second (\lgnmurms) moments of the \lgnmu distribution. There are two central features of the proposed procedure: a) a simplified model to describe the energy and mass evolution of \lgnmumean and \lgnmurms which minimizes the hadronic interaction model dependencies, and b) a comparison between the predictions of this model for a set of given composition scenarios and the data integrated in energy to maximize the discrimination power.

%%% secs 2, 3 and app

First in~\cref{sec:theory} we propose a simplified model to describe the energy and mass evolution of \lgnmumean and \lgnmurms. We argue that to a very good approximation only two parameters ($a$ and $b$) summarize all uncertainties of the currently used high energy hadronic interaction models. This simplification of the description of \lgnmumean and \lgnmurms with energy and mass is an important step in the analysis procedure because it minimizes the dependencies on hadronic interaction models in the interpretation of the data. In~\cref{sec:model} we use shower simulations to study the energy and mass evolution of \lgnmumean and \lgnmurms and to validate the model proposed in~\cref{sec:theory}. We also introduce in~\cref{sec:model} the algorithm developed to build the large set of simulations used in this paper. This simulation process is complemented in~\ref{sec:app}.

%%% sec 4

In~\cref{sec:moments} we introduce a set of six benchmark composition scenarios defined by the percentage of proton, helium, nitrogen and iron nuclei as a function of energy. Four composition scenarios are astrophysical motivated (based in Refs.~\cite{Berezinsky2005,Berezinsky2006,Allard2007,Biermann2012,Allard2015}) and two were derived from the \xmax measurements performed by the Pierre Auger Collaboration (based on Ref.~\cite{Aab2014a}). By using simulations we also study the energy evolution of \lgnmumean and \lgnmurms for each one of these scenarios and evaluate the effects of the uncertainties on the energy scale and on the absolute \nmu due to the misprediction by the hadronic interaction models.

%%% sec 5 and conclusions

In~\cref{sec:analysis} we show how the model proposed in~\cref{sec:theory} can be used to discriminate between these representative composition scenarios. The comparison of the model predictions for the composition scenarios with the data in an energy range is the important step of the analysis procedure proposed here because it maximizes the discrimination power allowing us to identify the most likely scenario that generated a set of \nmu data. This comparison is done by the traditional \cchi, which assumes the minimal value for the composition scenario which best describes the data. We use simulations to test our approach and show that it is possible to achieve a good discrimination between the chosen scenarios supposing a realistic case with the statistic to be collected during three years of data taking with the Pierre Auger Observatory Upgrade - AugerPrime. We also show that the systematic uncertainties in the energy reconstruction and on the absolute scale of the number of muons do not mix the composition scenarios. Hence we conclude in~\cref{sec:conclusion} that by using only the energy evolution of \lgnmumean and \lgnmurms it would be possible to identify, by comparing the composition scenarios to the data, the scenario which best describes the measurements of \nmu.

\section{A model for the energy and mass evolution of \lgnmu moments}
\label{sec:theory}

%=======================================================
% MEAN

In this section we present a model to describe the energy and primary mass evolution of the \lgnmu first and second moments. The Heitler-Matthews model~\cite{Matthews2005} is a semi-empirical description of the shower development which describes the dependencies of the mean \nmu as

\begin{equation}
\langle N_{\mu} \rangle_A = A^{1-\beta}N_{\mu}^p
\end{equation}

and

\begin{equation}
\langle N_{\mu} \rangle_E = \left( \frac{E}{\zeta^{\rm \pi}_c}\right)^{\beta} \; ,
\end{equation}

\noindent
where $N_{\mu}^p$ is the number of muons in a proton shower and $\zeta^{\rm \pi}_c$ is the pion critical energy, assumed to be equal to $20$ GeV in \cite{Matthews2005}. $\beta$ is often taken to be constant because its value is shown to vary in a small interval from 0.85 to 0.92~\cite{Matthews2005,Alvarez-Muniz2002}.

Both equations define a clear linear relation of \lgnmumean with energy and mass that can be summarized as

\begin{equation}
  \langle \log_{10} N_{\mu} \rangle_{E,A} = a + D_{A}\cdot \ln (A) + D_{E} \cdot \left( \log_{10}E - 19.0 \right),
  \label{eq:model:mean}
\end{equation}

\noindent
where $D_{E} = \beta \simeq 0.85-0.92$, $D_{A} = (1-\beta)\cdot \log_{10}e \simeq 0.434\cdot (1-\beta) \simeq 0.0347-0.0651$, and the energy $E$ is given in eV. Because of our lack of knowledge of the hadronic interactions at the highest energies, the value of $a$ is highly model dependent and presents a large variability. It can be written as $a = \log_{10} ( N_{\mu}^p ) - \beta \log_{10} ( \zeta^{\rm \pi}_c )$ and varies approximately from $6.5$ to $8.0$, depending on the hadronic interaction model. These values of $a$ were obtained using the simulations described in~\cref{sec:sim}.

%=======================================================
% RMS

In addition to the \lgnmumean, the \lgnmurms could also be modeled by the same approach. However, no analytic model has been proposed to describe the shower-to-shower fluctuations and our study relies on simulations to propose a similar description of \lgnmurms evolution with energy and mass. We propose that the \lgnmurms can be described as

\begin{equation}
  \sigma [\log_{10} N_{\mu} ]_{A} = \sigma [\log_{10} N_{\mu} ]_{\rm Fe} + b \cdot \left[\ln(A)-\ln(56)\right]^2 \; ,
  \label{eq:model:rms:lna}
\end{equation}

\noindent
where $ \sigma [\log_{10} N_{\mu} ]_{\rm Fe} $ is the \lgnmurms for iron nucleus initiated showers. Two main assumptions were used in this proposal: a) for a fixed primary (A), the \lgnmurms does not depend on energy and b) a quadratic dependency of \lgnmurms with \lna. These assumptions are justified in~\cref{sec:sim} via Monte Carlo simulation of the air shower.

The description of the \lgnmurms is analogous to the deduction of \lgnmumean using the Heitler-Matthews models in the following way. We will show in~\cref{sec:sim} that, for the purposes of this paper's analysis, $\sigma [\log_{10} N_{\mu} ]_{\rm Fe} $ can be taken to be constant, in other words, the small model dependence of $\sigma [\log_{10} N_{\mu} ]_{\rm Fe} $ can be ignored. On the other hand, $b$ changes significantly with the hadronic interaction model, which reflects the theoretical uncertainties concerning the muonic component description.

\cref{eq:model:mean} and~\cref{eq:model:rms:lna} summarize the first step of this paper. These equations offer a simple, but good description of the two first moments of the \lgnmu distribution with energy and mass. The uncertainties due to hadronic interaction model descriptions are only significant for two parameters, $a$ and $b$, while for the further parameters there is a good agreement between their predictions. The quality of the description given by~\cref{eq:model:mean} and~\cref{eq:model:rms:lna} is going to be numerically studied in the next section.

%=======================================================
% Mixture of primaries

For a mixture of primaries in which each primary particle type, $i$,  has mass $A_i$ and contributes to the total flux with a fraction given by $f_i$, we can show that \lgnmumean and \lgnmurms of the mixture ($mix$) can be calculated as follows

\begin{equation}
  \begin{split}
    \langle \log_{10} N_{\mu} \rangle_{\rm mix} &= \sum_{i}f_i \cdot \langle \log_{10} N_{\mu} \rangle_{A_i} \\
    & = a + D_{A}\cdot \langle \ln (A) \rangle_{\rm mix} + D_{E} \cdot \left( \log_{10}E - 19.0 \right),
  \end{split}
  \label{eq:model:mean:tot}
\end{equation}

\noindent
and
\begin{equation}
  \begin{split}
   & \sigma^2[\log_{10} N_{\mu} ]_{\rm mix} = \\
   &=\sum_i f_i \cdot \left[  \left( \langle \log_{10} N_{\mu} \rangle_{A_i}- \langle \log_{10} N_{\mu} \rangle_{\rm mix} \right)^2 +  \sigma^2[\log_{10} N_{\mu}]_{A_i} \right]\;.
  \end{split}
  \label{eq:model:rms:sum}
\end{equation}

\noindent
Using~\cref{eq:model:mean} we can write

\begin{equation}
  \begin{split}
   & \sigma^2[\log_{10} N_{\mu} ]_{\rm mix} = \\
   &=\sum_i f_i \cdot \left[ D^2_{A}\cdot \left( \ln(A_{i}) - \langle \ln(A) \rangle_{\rm mix} \right)^2 +  \sigma^2[\log_{10} N_{\mu}]_{A_i} \right] \; .
  \end{split}
  \label{eq:model:rms:sum}
\end{equation}

\noindent
Note that $\sigma^2[\log_{10} N_{\mu} ]_{\rm mix}$ does not depend on $a$. The dependence on $b$ is implicit in the $\sigma^2[\log_{10} N_{\mu}]_{A_i}$ term.

%%%%%%%%%%======================================%%%%%%%%%%%%%%%%%%

\section{Simulation studies of \lgnmu moments}
\label{sec:sim}

In this section we briefly describe the procedure adopted to produce simulated \lgnmu distributions that are extensively employed in the following sections of this paper. The present discussion is complemented by~\ref{sec:app} where more details about the simulations are given. Furthermore, in this section we also use the simulated showers to validate the \lgnmu moment descriptions proposed in~\cref{sec:theory} and to study the energy evolution of \lgnmu moments for a set of mass composition scenarios.

\subsection{Simulation technique}
\label{sec:model}

In our analysis we aim to assess the number of muons measured in UHECR experiments. A combination of detector technology, observatory altitude, spatial configuration of the detectors and analysis procedures determines the lateral distance range and the energy threshold of detectable muons. To avoid saturation of the detectors (close to the shower axis) and large statistical fluctuations (far from the shower axis), a fiducial lateral distance range is commonly defined to get the lateral distance function integrated. Therefore, the measured number of muons  (\nmuexp) is not the total number of muons at the ground but only a sample of them above an energy threshold and within a distance range.

In this paper \nmuexp is defined as the number of muons with energy above $0.2$ GeV reaching the ground (1400 m above sea level, the Auger mean altitude) at a distance between $500$ m and $2000$ m from the shower axis. This choice is motivated by the design of the main current high energy cosmic ray experiments, for example, the Pierre Auger Observatory~\cite{Auger2015} and Telescope Array~\cite{Kawai2008}.

The muons spatial and energy distributions at the ground can be evaluated by CORSIKA~\cite{Heck1998a} (version 7.4000), which is a full Monte Carlo code able to perform 3D shower simulations. \nmuexp could be determined by CORSIKA, in despite of its high computational cost~\cite{Ortiz2005}. CONEX~\cite{Bergmann2007} (version 2r4.37) is a very fast hybrid simulation code which combines full Monte Carlo with solutions of one-dimensional cascade equations. From CONEX simulations it is possible to determine the total number of muons at the ground above $1$ GeV (\nmutotal).

\nmutotal and \nmuexp can be simultaneously obtained from full simulated showers (CORSIKA), allowing us to parametrize the relation between them. We propose the following parametrization:

\begin{equation}
  N_{\mu}^{\rm meas} = R(E,X_{\rm max}) \cdot N_{\mu}^{\rm tot},
\end{equation}

\noindent
where the conversion factor $R$ should be determined for each primary and depends on the energy and \xmax. The parametrization of $R(E,X_{\rm max})$ is explored in detail in~\ref{sec:app}. The \xmax dependence of the factor $R$ ensures that the parametrization takes into account the shower-to-shower fluctuations due to the variance of the first interaction depth. Furthermore, the most relevant physical processes responsible for muons production in showers are reliably reproduced by the CONEX simulations, and consequently they should also be represented in \nmuexp. As shown in~\ref{sec:app}, the \nmuexp distributions obtained based on the proposed parametrization are in very good agreement with the ones obtained from full Monte Carlo simulation.

The parametrization was done only for shower at 38$^\circ$ zenith angle. The zenith angle dependence can be taken into account by simulating other primaries with the corresponding arrival direction and by dividing the data in zenith angle intervals.

\subsection{Simulating \lgnmu moments}

We generated 60000 CONEX (version 2r4.37) showers with energies between $10^{18.4}$ and $10^{19.6}$ eV, for four primaries (proton, helium, nitrogen and iron) and two hadronic interaction models (EPOS-LHC~\cite{Werner2007} and QGSJetII-04~\cite{Ostapchenko2010}). The showers are distributed uniformly in \lge and the zenith angle is fixed at 38$^{\circ}$. From the $R(E,X_{\rm max})$ parametrization of~\ref{sec:app}, the CONEX showers were converted into a set of \nmuexp. %In the following, \nmuexp is denoted only by \nmu, for simplicity.

\begin{figure*}
  \subfloat[]{
    \includegraphics[width=0.47\textwidth]{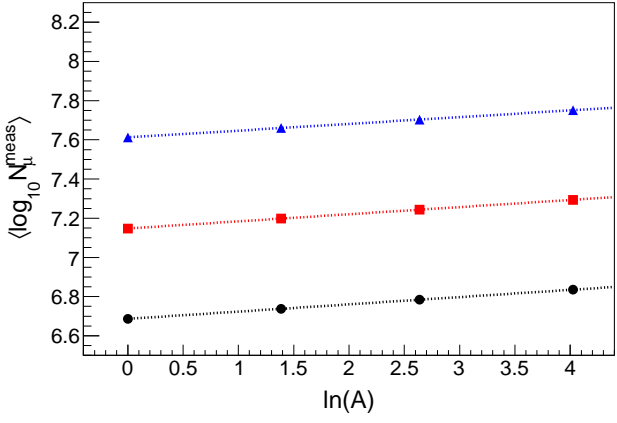}
  }
  \subfloat[]{
    \includegraphics[width=0.47\textwidth]{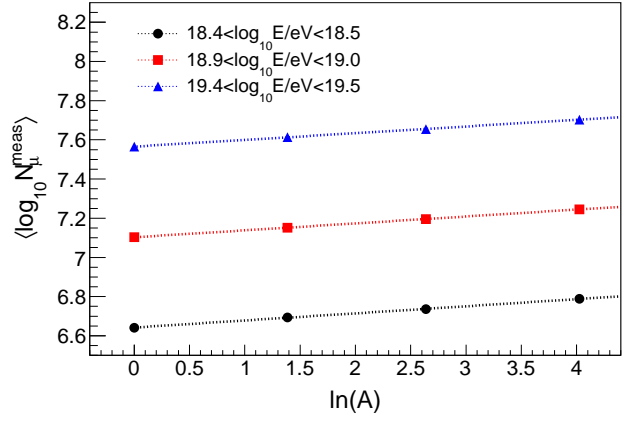}
  }

  \caption{\lgnmumeanexp as a function of \lna for both hadronic interaction models, (a) EPOS-LHC and (b) QGSJetII-04 and three energy intervals. The dotted lines are the results of the linear fit, represented in~\cref{eq:model:mean}. The statistical error bars are smaller than the markers.}
  \label{fig:model:lna:mean}
\end{figure*}

\cref{fig:model:lna:mean} shows the evolution of \lgnmumeanexp with the primary mass for three energy intervals. Lines are the result of a linear fit which demonstrates the dependence of \lgnmumeanexp with mass as proposed in~\cref{eq:model:mean}. The fits resulted in $D_{A} \simeq 0.034-0.037$ and $a \simeq 6.64-7.70$, with errors from the fit less than $0.0005$ and $0.005$, respectively.

\begin{figure*}
  \subfloat[]{
    \includegraphics[width=0.47\textwidth]{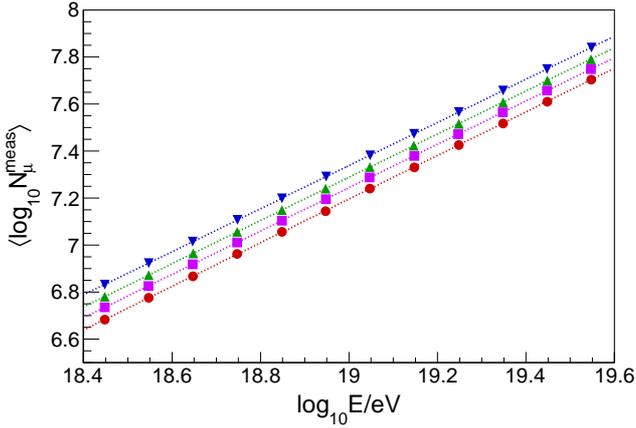}
  }
  \subfloat[]{
    \includegraphics[width=0.47\textwidth]{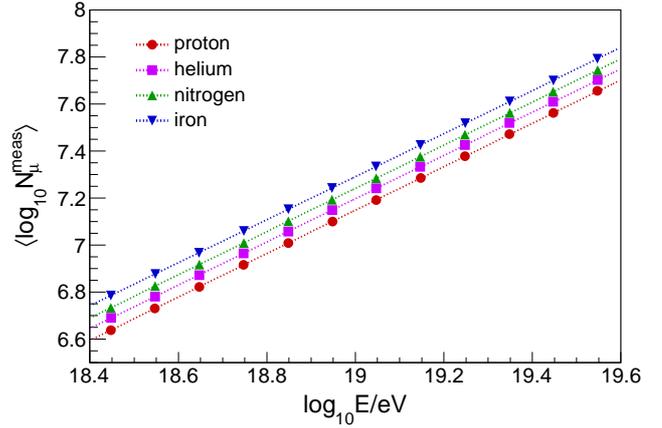}
  }

  \caption{\lgnmumeanexp as a function of \lge for both hadronic interaction models, (a) EPOS-LHC and (b) QGSJetII-04, and four primaries (proton, helium, nitrogen and iron). The dotted lines are the results of the linear fit, represented in~\cref{eq:model:mean}. The statistical error bars are smaller than the markers.}
  \label{fig:model:lge:mean}
\end{figure*}

\begin{figure*}
  \subfloat[]{
    \includegraphics[width=0.47\textwidth]{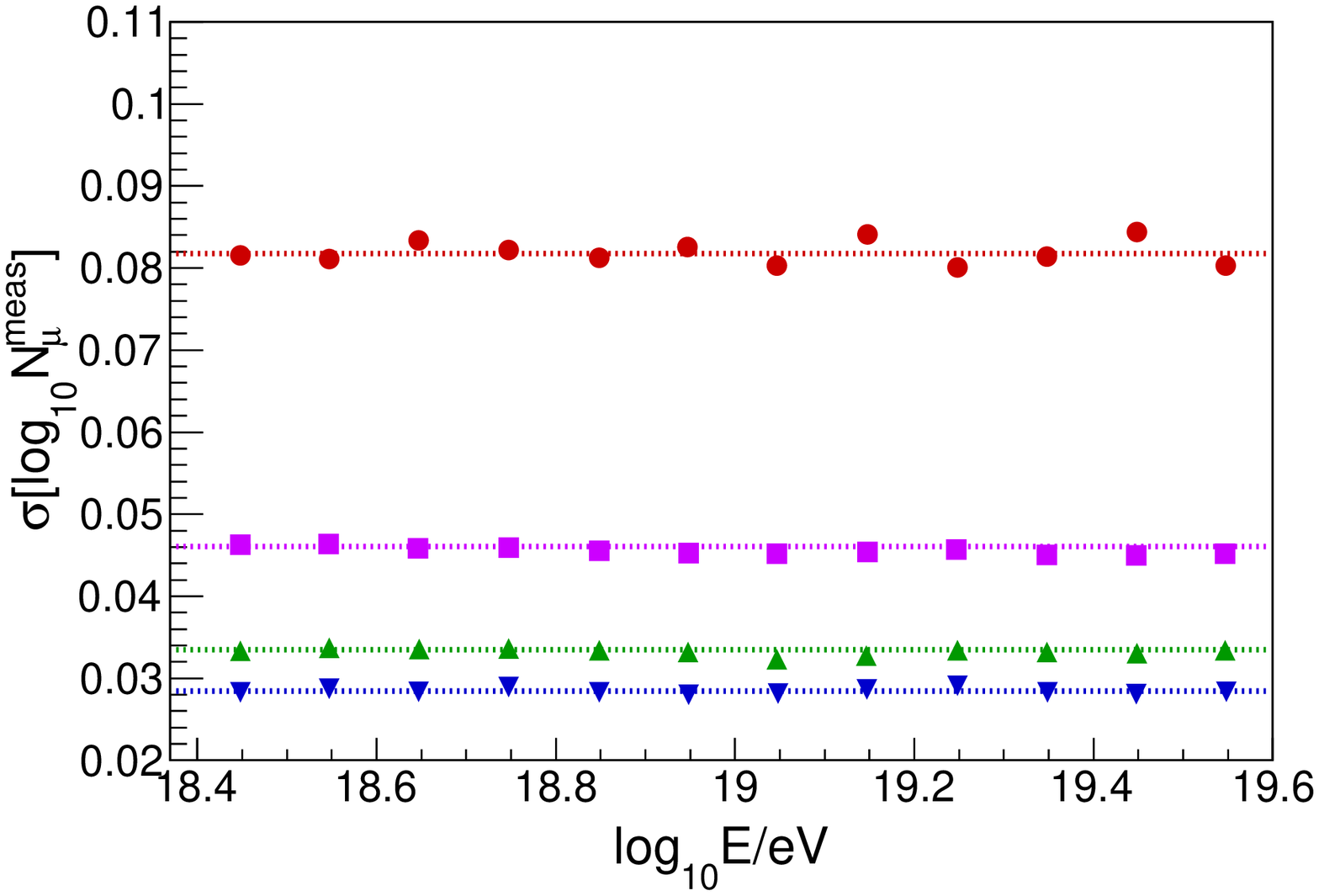}
  }
  \subfloat[]{
    \includegraphics[width=0.47\textwidth]{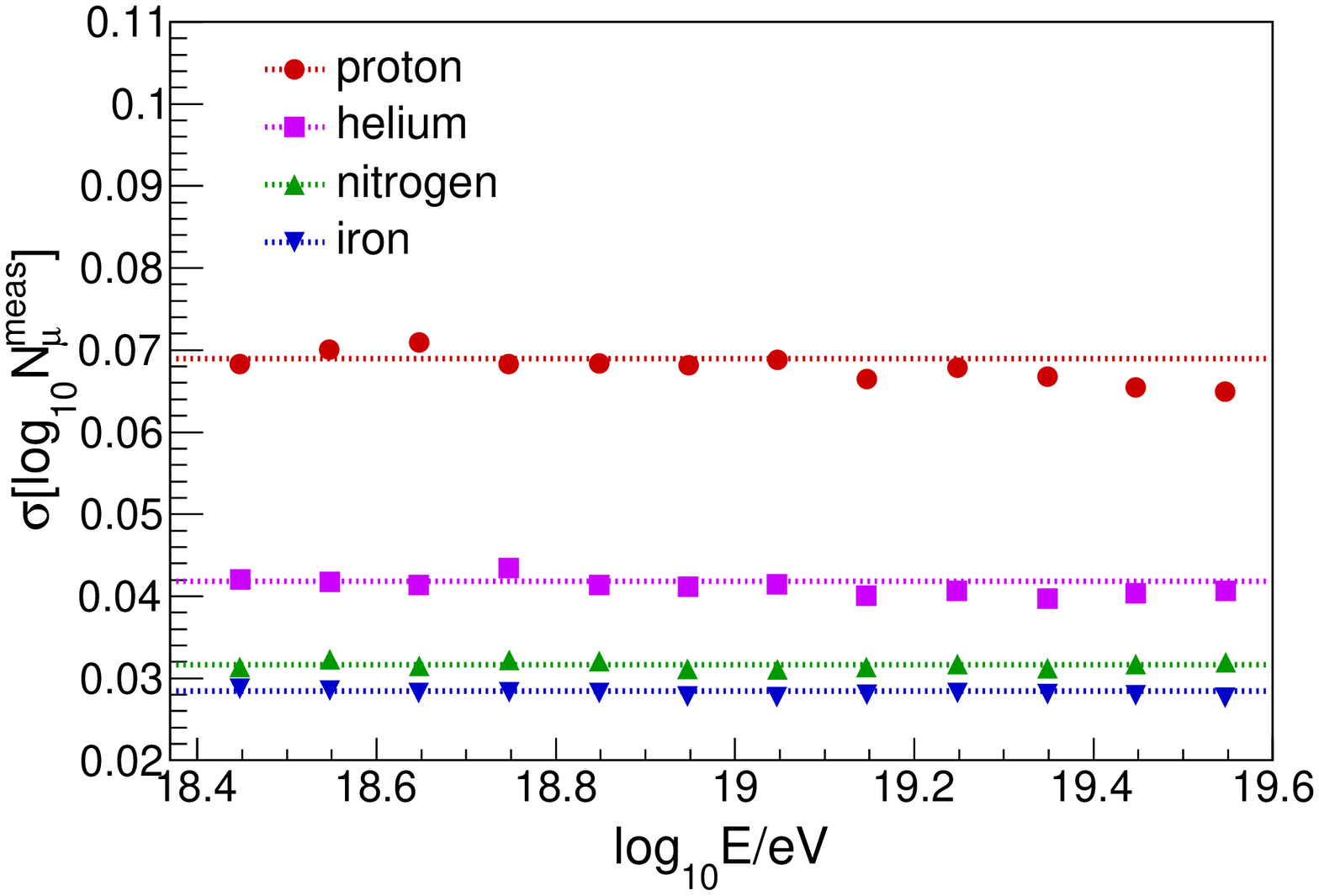}
  }

  \caption{\lgnmurmsexp as a function of \lge for both hadronic interaction models, (a) EPOS-LHC and (b) QGSJetII-04, and four primaries (proton, helium, nitrogen and iron). The dotted lines are the results of the fit of a constant energy function. The statistical error bars are smaller than the markers.}
  \label{fig:model:lge:rms}
\end{figure*}

The energy evolution of \lgnmumeanexp is shown in~\cref{fig:model:lge:mean}, where one can note the linear behavior as proposed in~\cref{eq:model:mean}. The fits resulted in $D_{E} \simeq 0.915-0.928$, with errors from the fit less than $0.0003$. The energy evolution of \lgnmurmsexp is shown in~\cref{fig:model:lge:rms}. Note the flatness of \lgnmurmsexp and the coincidence of the \lgnmurmsexp constant value of iron initiated showers for both hadronic interaction models. These figures validate both assumptions made in~\cref{sec:theory} concerning \lgnmurmsexp.

%=====================================================================
%simulation: mean log nmu versus mass
%=====================================================================
%simulation: mean log nmu versus energy

\begin{figure*}
  \subfloat[]{
    \includegraphics[width=0.47\textwidth]{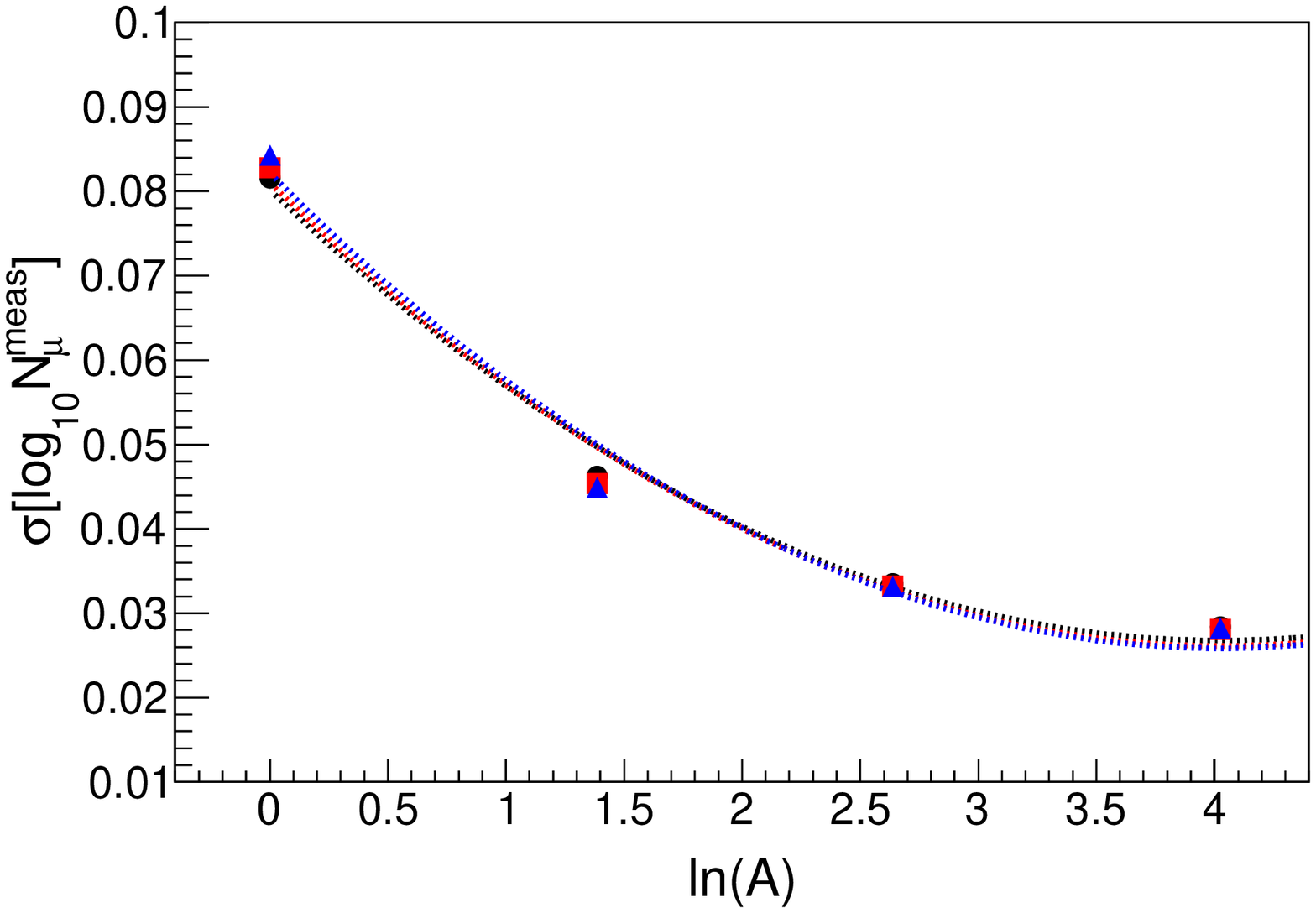}
  }
  \subfloat[]{
    \includegraphics[width=0.47\textwidth]{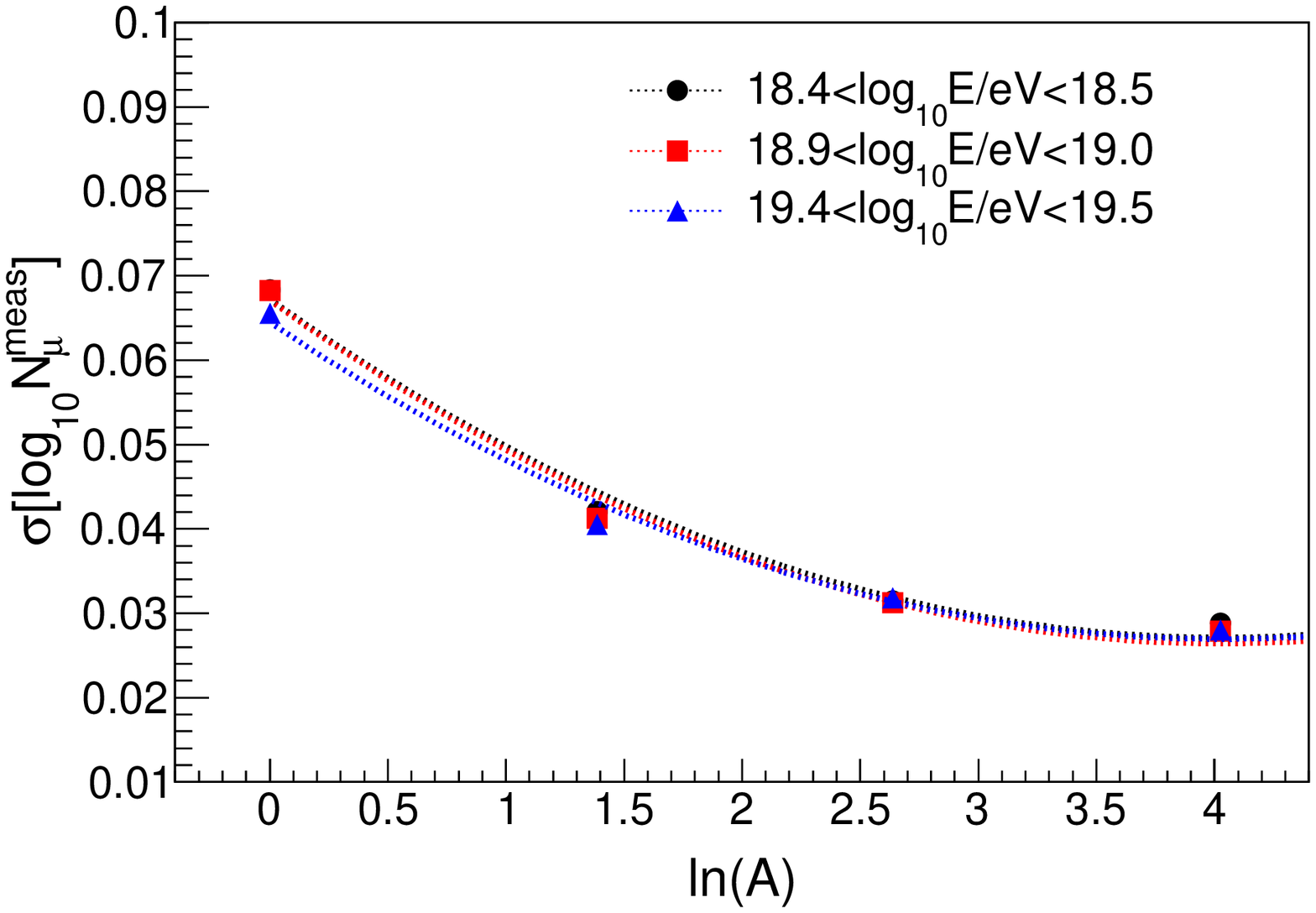}
  }

  \caption{\lgnmurmsexp as a function of \lna for both hadronic interaction models, (a) EPOS-LHC and (b) QGSJetII-04, and three energy intervals. The dotted lines are the results of the fit of~\cref{eq:model:rms:lna}. The statistical error bars are smaller than the markers.}
  \label{fig:model:lna:rms}
\end{figure*}

The primary mass dependence of \lgnmurmsexp can be seen in~\cref{fig:model:lna:rms} for three energy intervals and both hadronic interaction models. The dashed lines are the quadratic curves shown in~\cref{eq:model:rms:lna} fitted to the points. The fits resulted in $ \sigma [\log_{10} N_{\mu} ]_{\rm Fe} $ being indeed nearly constant, varying from $0.0258$ to $0.0275$, with errors from the fits less than $0.003$. The fit also resulted in $b = 0.0024 \pm 0.0002$ for QGSJetII-04 and $b = 0.0033 \pm 0.0003$ for EPOS-LHC. The simulations shown in this section confirmed all the assumptions made in~\cref{sec:theory}.

%=====================================================================
%simulation: rms log nmu versus mass

\section{Mass composition scenarios and the energy evolution of the \lgnmu moments}
\label{sec:moments}

In this section we simulate the energy evolution of \lgnmuexp moments for six mass composition scenarios, which are defined by setting the fractions $f_i(E)$ of the total flux corresponding to each particle with mass $A_i$. Given $f_i(E)$ and $A_i$ we can calculate \lgnmumeanexp and \lgnmurmsexp as a function of energy using the procedure described in~\cref{sec:theory,sec:sim}.

The mass composition scenarios we used are divided in two groups. The first one includes the astrophysical motivated scenarios, which are labeled by the letter A. The second group includes two scenarios obtained from the \xmax distributions fit performed by the Pierre Auger Collaboration \cite{Aab2014,Aab2014a} and they are labeled by the letter X. Below, we present a brief description of the composition scenarios, which can be skipped by the reader that is familiar with the subject.

\begin{description}

\item \underline{Scenario A1}: This scenario proposes a pure proton flux. It was the first model proposed to explain the dip in the energy spectrum as the effect of pair-production in the propagation of the UHECR. This model was originally proposed in Refs.~\cite{Allard2005a,Berezinsky2005} and was also discussed in Refs.~\cite{Berezinsky2006,Allard2007}(labeled as Model B in Ref.~\cite{Allard2007}).

\item \underline{Scenario A2}: This scenario assumes a mixed source composition with abundances similar to the data at lower energies. It was proposed by Allard et al. (labeled as Model A in Ref.~\cite{Allard2007}). In this model the ankle is explained as the transition in the predominance of the flux from the galactic to the extra-galactic component. The abundances are originally given for five groups of nuclei, however, in this paper the fluxes of the two heaviest groups were summed into the iron component.

\item \underline{Scenario A3}: Biermann \& de Souza~\cite{Biermann2012} have proposed a model in which the observed cosmic ray energy spectrum from $10^{15.0}$ to $3\times 10^{20.0}$ eV is explained by the galactic and only one extra-galactic source, the radio galaxy Cen A. In this model the element abundances from extra-galactic origin are similar to the galactic ones, but shifted up in energy because of the relativistic shock in the jet emanating from the active black hole. The abundances are originally given for six groups of nuclei, however, in this paper the flux of the element group Ne-S was summed into the nitrogen flux and the flux of the Cl-Mn group was summed into the iron group flux.

\item \underline{Scenario A4}: The model proposed by Globus et al.~\cite{Allard2015} describes the whole cosmic ray spectrum by superposing a rigidity dependent galactic component and a generic extra-galactic component. This model gives an adequate description of the energy spectrum and the moments of the \xmax distribution measured by the Pierre Auger Observatory.

\item \underline{Scenario X1}: It has been shown by the Pierre Auger Collaboration that the measured \xmax distributions can be well described by a combination of four components~\cite{Aab2014,Aab2014a}. By fitting the \xmax simulated distributions to the data, the abundances of the separate components were obtained as a function of energy. This scenario is based on the abundances obtained by using the hadronic interaction model QGSJetII-04. However, the abundances obtained with Sibyll2.1 are also very close to the one we used. In order to minimize point-to-point fluctuations, we used here a smooth curve fitted to the fractions obtained in the Auger analysis~\cite{Aab2014a}.

\item \underline{Scenario X2}: This scenario was obtained by fitting \xmax distributions measured by the Pierre Auger Observatory using showers simulated with the EPOS-LHC hadronic interaction model. The procedure is the same as the one adopted for Scenario X1.

\end{description}

\begin{figure}
  \centering
  \includegraphics[width=0.47\textwidth]{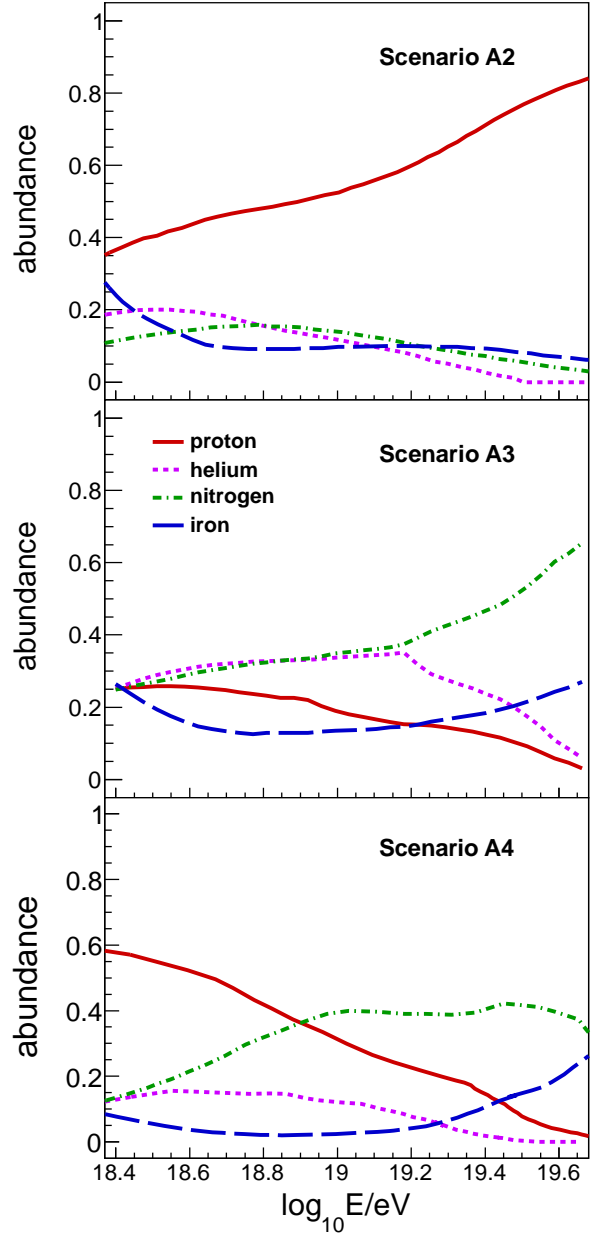}

  \caption{The composition component abundances as a function of \lge for the mass composition scenarios A2, A3 and A4 (see text).}
  \label{fig:scenarios:astro}
\end{figure}

\begin{figure}
  \centering
  \includegraphics[width=0.47\textwidth]{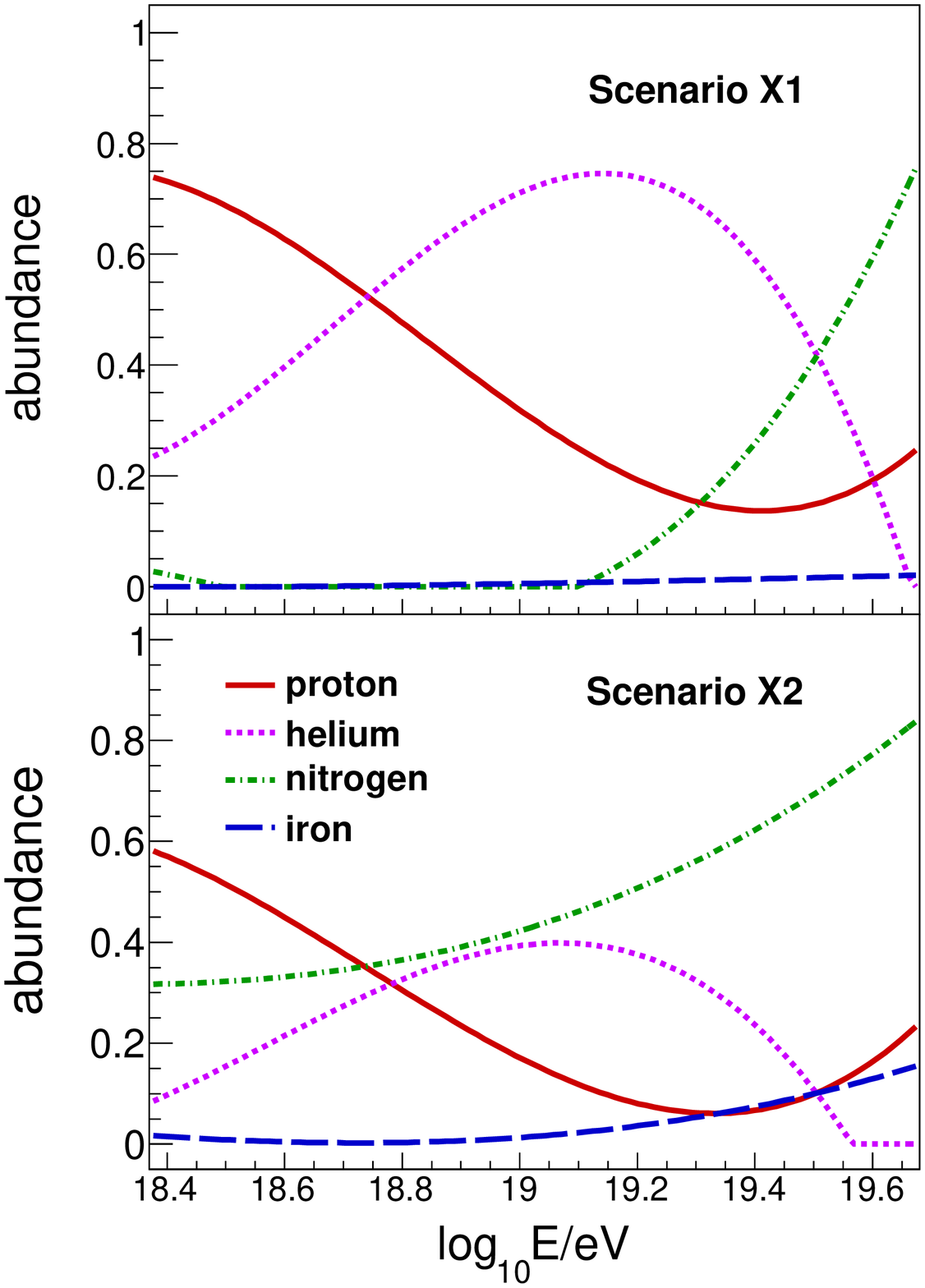}

  \caption{The composition component abundances as a function of \lge for the mass composition scenarios X1 and X2 (see text).}
  \label{fig:scenarios:xmax}
\end{figure}

The merging of components done for models A2 and A3 is necessary to allow us to use the parametrization elaborated in~\cref{sec:sim}. Since we present in this paper only the analysis procedure, verified with simulations, this choice has no limiting consequence. Besides that, the systematic uncertainties of the abundances obtained from the scenarios are also going to be neglected here. \cref{fig:scenarios:astro,fig:scenarios:xmax} show the abundances for each scenario in the energy range from \energy{18.4} to \energyEV{19.6} as explained above. Scenario A1 is not shown because it assumes a 100\% proton flux.

%=====================================================================
%comp scenarios

% FIGURES SIDE BY SIDE FOR ARXIV VERSION
%\begin{figure}[h]
%  \centering
%  \begin{minipage}[b]{0.47\textwidth}
%    \centering
%    \includegraphics[width=0.98\textwidth]{figures/scenarios_astro}
%
%    \caption{The composition component abundances as a function of \lge for the mass composition scenarios A2, A3 and A4 (see text).}
%    \label{fig:scenarios:astro}
%  \end{minipage}%
%  \hspace{0.5cm}
%  \begin{minipage}[b]{0.47\textwidth}
%    \centering
%    \includegraphics[width=0.98\textwidth]{figures/scenarios_xmax}

%    \caption{The composition component abundances as a function of \lge for the mass composition scenarios X1 and X2 (see text).}
%    \label{fig:scenarios:xmax}
%  \end{minipage}
%\end{figure}

\begin{figure*}
  \subfloat[]{
    \includegraphics[width=0.47\textwidth]{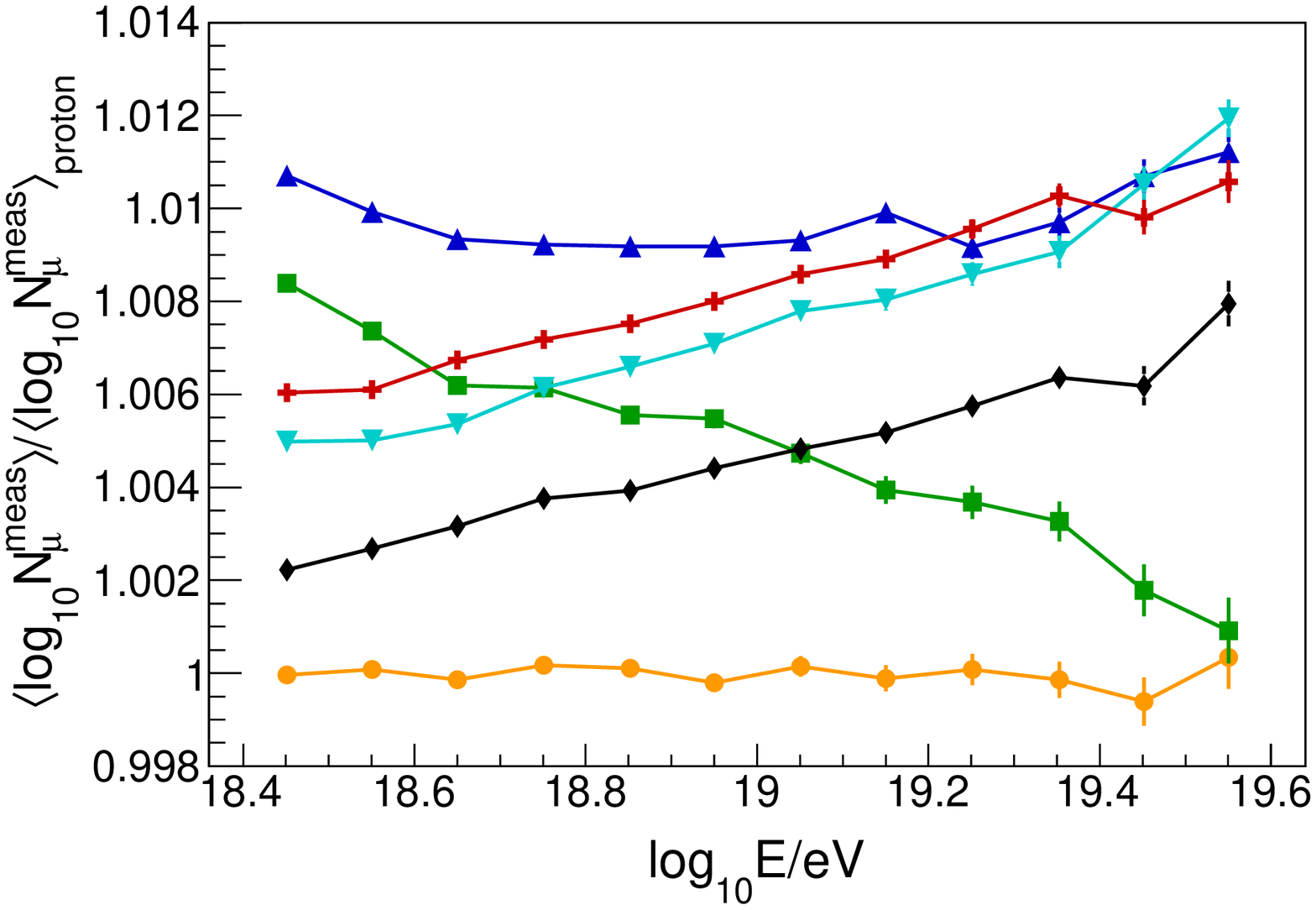}
    \label{fig:moments:mean}
  }
  \subfloat[]{
    \includegraphics[width=0.47\textwidth]{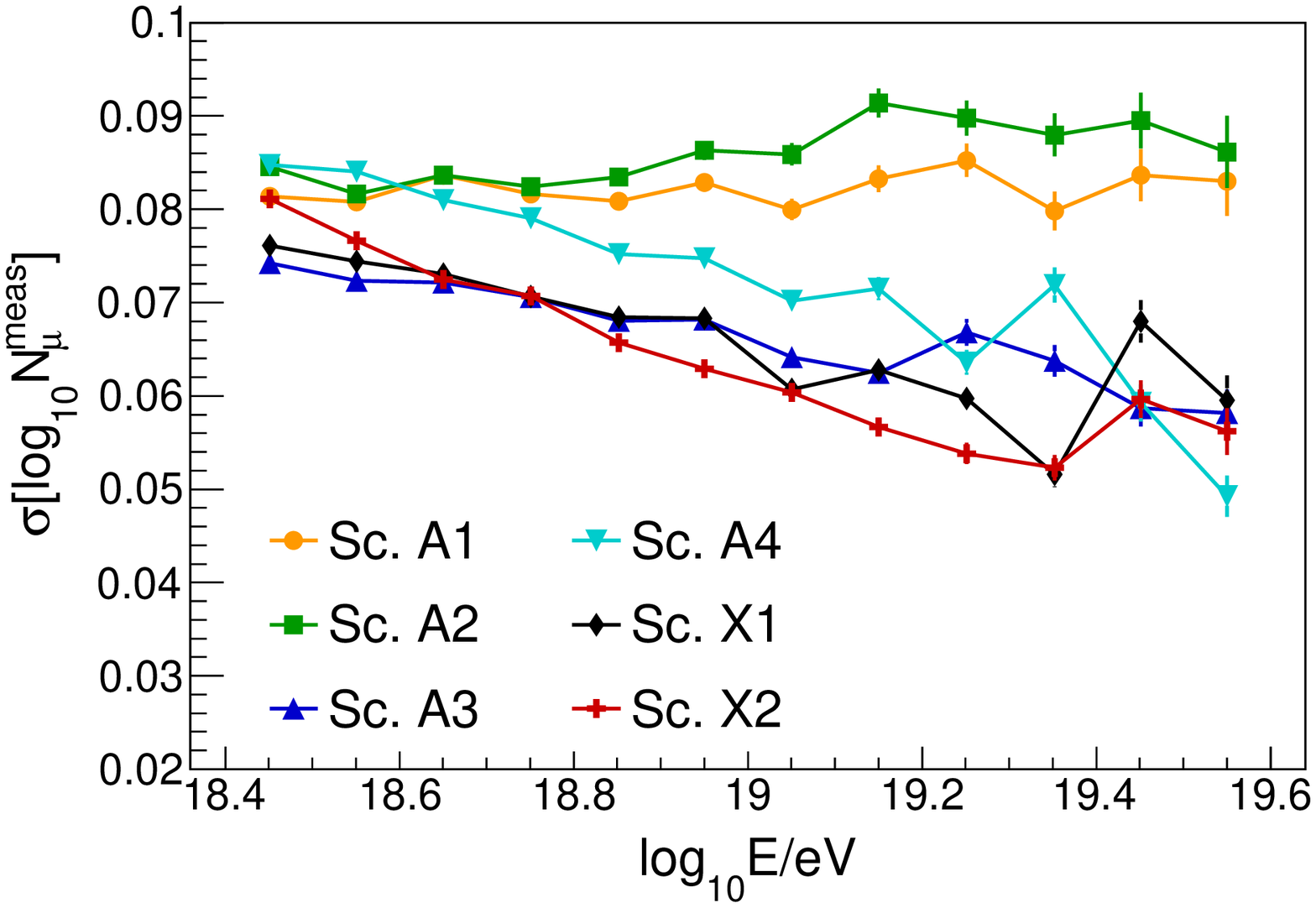}
    \label{fig:moments:rms}
  }

  \caption{Energy evolution of (a) \lgnmumeanexp and (b) \lgnmurmsexp for the six composition scenarios described in the text. The values of \lgnmumeanexp are divided by the corresponding value of pure proton composition for better visualization. The hadronic interaction model used was EPOS-LHC. }
  \label{fig:moments}
\end{figure*}

\cref{fig:moments} shows the energy evolution of the \lgnmumeanexp and \lgnmurmsexp for all mass composition scenarios. The error bars correspond to the one sigma fluctuation of the mean value considering the statistics from three years of AugerPrime data (3000 km\expo{2} of muon detectors). The all particle flux was taken from Ref.~\cite{AugerSpec2013}. \cref{fig:moments:mean} shows the mean normalized to the proton simulation for better visualization.

%=====================================================================
%simulation: mixture of primaries: mean and rms log nmu versus energy

%=====================================================================
% simulation: systematics effects
\begin{figure*}
  \subfloat[]{
    \includegraphics[width=0.47\textwidth]{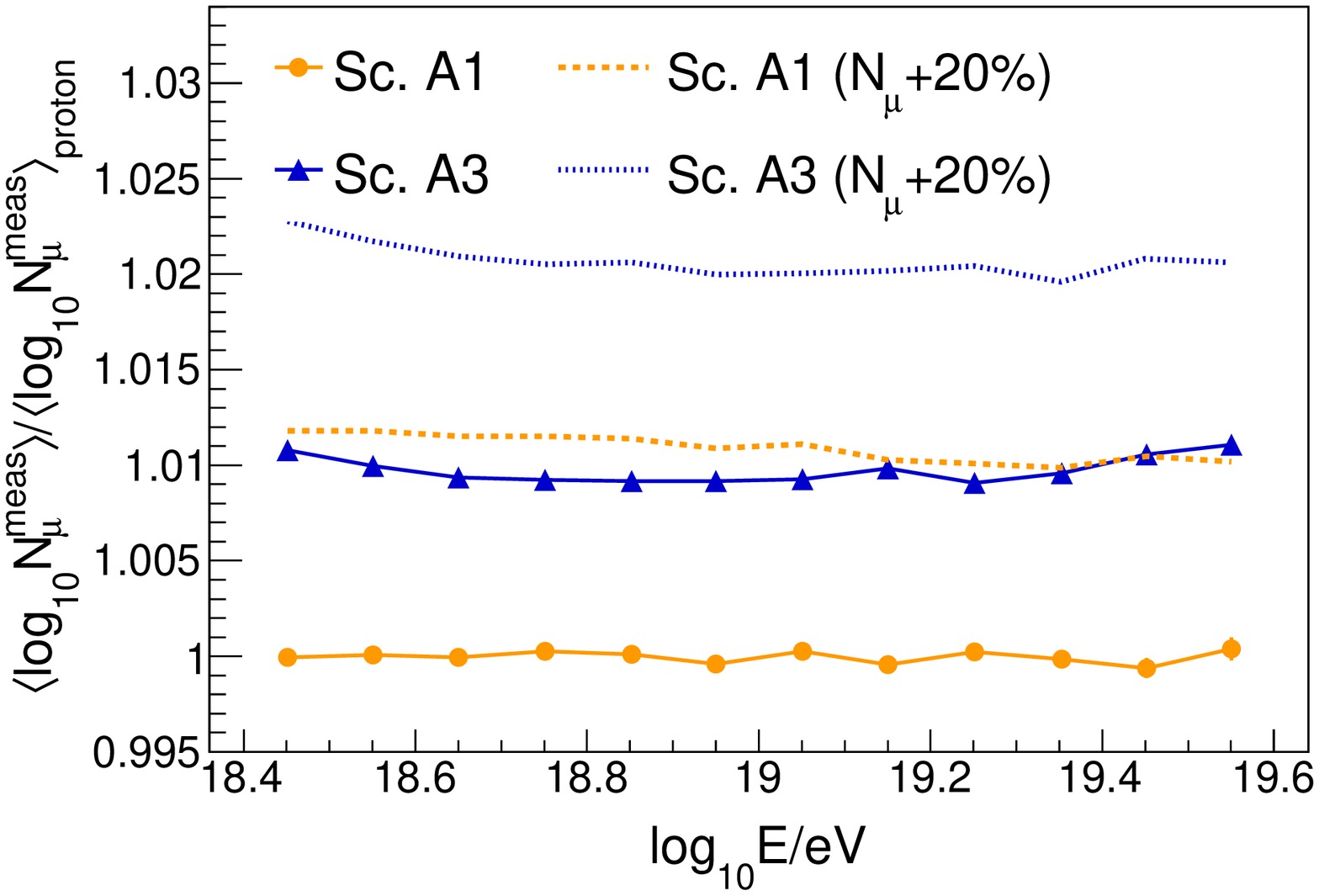}
    \label{fig:moments:syst:nmu}
  }
  \subfloat[]{
    \includegraphics[width=0.47\textwidth]{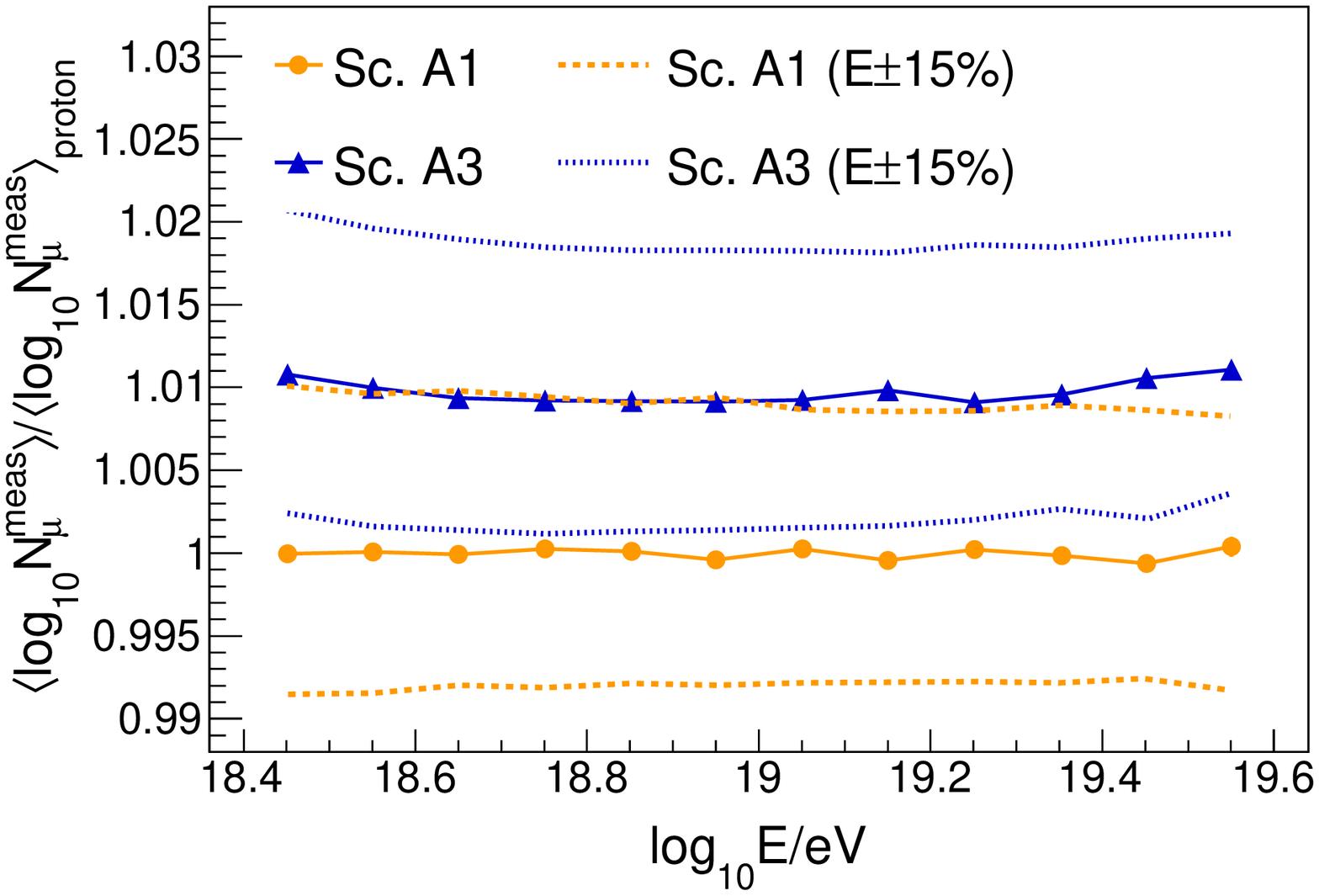}
    \label{fig:moments:syst:en}
  }
  \caption{Energy evolution of \lgnmumeanexp for 2 mass composition scenarios, A1 and A3. The dashed lines show the effects of (a) an increase of 20\% in the \nmu and (b) a variation of $\pm$15\% in energy. The values of \lgnmumeanexp are divided by the corresponding value of pure proton composition for better visualization. The hadronic interaction model used was EPOS-LHC.}
  \label{fig:moments:syst}
\end{figure*}

%%%%%%%%%%%%%%%%%%====================================%%%%%%%%%%%%%%%%%%%%%%%%%%%%%%%

\section{Discrimination between mass composition scenarios}
\label{sec:analysis}

Given the theoretical uncertainties on the \nmu predictions and the systematic uncertainties on the energy reconstruction, the question we would like to answer in this section is how it is possible to discriminate between the mass composition scenarios shown above using the evolution of the \lgnmuexp moments with energy. Examining~\cref{fig:moments} it might seem easy to differentiate the scenarios by using the absolute value or the evolution of the \lgnmuexp moments with energy. However, if we include in this figure the uncertainties in the hadronic interaction model and systematic in energy reconstruction the interpretation of the data is not straightforward.

\cref{fig:moments:syst} shows how the uncertainties on the hadronic interaction model predictions and on the energy reconstruction influence the interpretation of the \lgnmumeanexp in terms of composition. We show in this figure the extreme composition scenarios (A1 and A3), since the other four scenarios lie within them. In \cref{fig:moments:syst:nmu} we calculate \lgnmumeanexp for scenarios A1 and A3 adding arbitrarily 20\% more muons to the simulation predictions to mimic the theoretical uncertainties in the hadronic interaction model predictions~\cite{Aab2015a}. Even the extreme models A1 and A3 would overlap if the uncertainty is considered. In \cref{fig:moments:syst:en} we calculate \lgnmumeanexp for scenarios A1 and A3 and changed the simulated energy by $\pm 15\%$ in order to evaluate the effect of the systematic uncertainty in the energy reconstruction. Once more it is clear that even the extreme scenario cases cannot be distinguished anymore. Moreover a combination of both uncertainties in the \nmu predictions and energy applied to this analysis would make the discrimination between the scenarios even harder. The conclusion is clear: the measurement of \nmu does not lead to a straightforward interpretation of the data in terms of composition if all the uncertainties are considered.

It is worthwhile to remember here how the interpretation of the \xmax measurement is done. The Pierre Auger Collaboration, for example,  fits $f_i$ to the measured \xmax distribution in bins of energy~\cite{Aab2014a}. The calculation of $f_i$ depends on simulation and therefore on the hadronic interaction model. However, because the electromagnetic cascade of the shower dominates the determination of the \xmax position, the discrepancy between the hadronic interaction model \xmax predictions is minimized. The difference in \xmaxmean is at most $20$ g/cm$^2$ and in \xmaxrms is $6$ g/cm$^2$ for the most often used hadronic interaction models (EPOS-LHC, Sibyll2.1 and QGSJetII-04)~\cite{Aab2014}. Given the small differences in the predictions of \xmax and its consistency with data, the fit of $f_i$ leads to acceptable differences in the calculation of $f_i$ for different hadronic interaction models and then to mass composition scenarios which are physically consistent.

Unfortunately, the same procedure cannot be applied to \nmu because of the discrepancies between the hadronic interaction model predictions and the inconsistency between simulations and data. It is known that the simulations are off by at least 20\% in the calculation of \nmu~\cite{Farrar2013,Aab2015a}. A fit of $f_i$ based on the \nmu distribution would lead to non-physical results. Therefore we propose an alternative analysis to discriminate between composition scenarios. The idea is to fix $f_i$, choosing a mass composition scenario, and fit the data with the energy evolution of \lgnmuexp moments to search for the scenarios which better describe the data.

If the composition ($f_i$) were known by an independent measurement, this procedure would allow us to calculate $a$ and $b$ and constrain the hadronic interaction models by limiting fundamental properties of the interactions. This hypothesis needs to be explored further by using the results from the \xmax measurement to fix $f_i$.

We propose here a procedure that allows a statistically robust test of composition scenarios against data. The method starts by using the model proposed in~\cref{sec:theory} to predict the energy evolution of \lgnmumeanexp and \lgnmurmsexp for a given composition scenario. Here, all the parameters of the model are fixed, except $a$ and $b$. The next step is to compare these predictions with data and find the values of $a$ and $b$ which make the model most similar to the data. This can be done by a \cchi minimization. The minimal values of \cchi determine which scenario best describes the data. Since $a$ and $b$ take all the hadronic interaction model dependence, the composition scenario can be tested independently of hadronic interaction model limitations.

We explored this analysis proposal by choosing a composition scenario as if it would represent the true measurement, and we name it \textit{true scenario}. We generate the  \lgnmu moments as a function of energy for the \textit{true scenario} using the simulation described in~\cref{sec:sim} in order to emulate the real data. Here, the energy bins are defined by $12$ intervals of  width $\Delta \log_{10}(E/{\rm eV}) = 0.1$, from $10^{18.4}$ to $10^{19.6}$ eV. This choice is mainly motivated by the Auger experimental acceptance, which reaches a 100\% efficient trigger probability around $10^{18.4}$ eV~\cite{Abraham2010d}. The number of events in each bin is determined by considering three years of data taken by the full array of Auger, following the energy spectrum of Ref.\cite{AugerSpec2013}.

It is important to note that the simulations used here do not take into account any detector effects or zenith angle dependence. Although in this paper we do not intend to approach these issues because the focus here are on the general aspects of the analysis, it is clear that in practical applications of the method one should deal with these experimental difficulties. The detector effects, like resolution and limited acceptance, could be addressed by unfolding or unbiasing techniques once the detector response is well known. One example of these process is the Auger analysis of \xmax moments~\cite{Aab2014}. The zenith angle dependence could be addressed in a conservative approach by dividing the data in zenith angle intervals or by correcting the data using a \textit{constant intensity cut} (CIC) method~\cite{Hersil1961,Edge1973,Nagano1984}. This later class of method has been successfully used, for example, to determine the shower size parameter by Pierre Auger~\cite{DiGiulio2009} and KASCADE-Grande Collaboration~\cite{Apel2016} and to correct the \nmu parameter by KASCADE-Grande Collaboration~\cite{Arteaga-Velazquez2009}. The systematics uncertainties from these procedure are usually small ($<10\%$) and should be taken into account in a realistic approach of our method.

\begin{figure*}
  \subfloat[]{
    \includegraphics[width=0.47\textwidth]{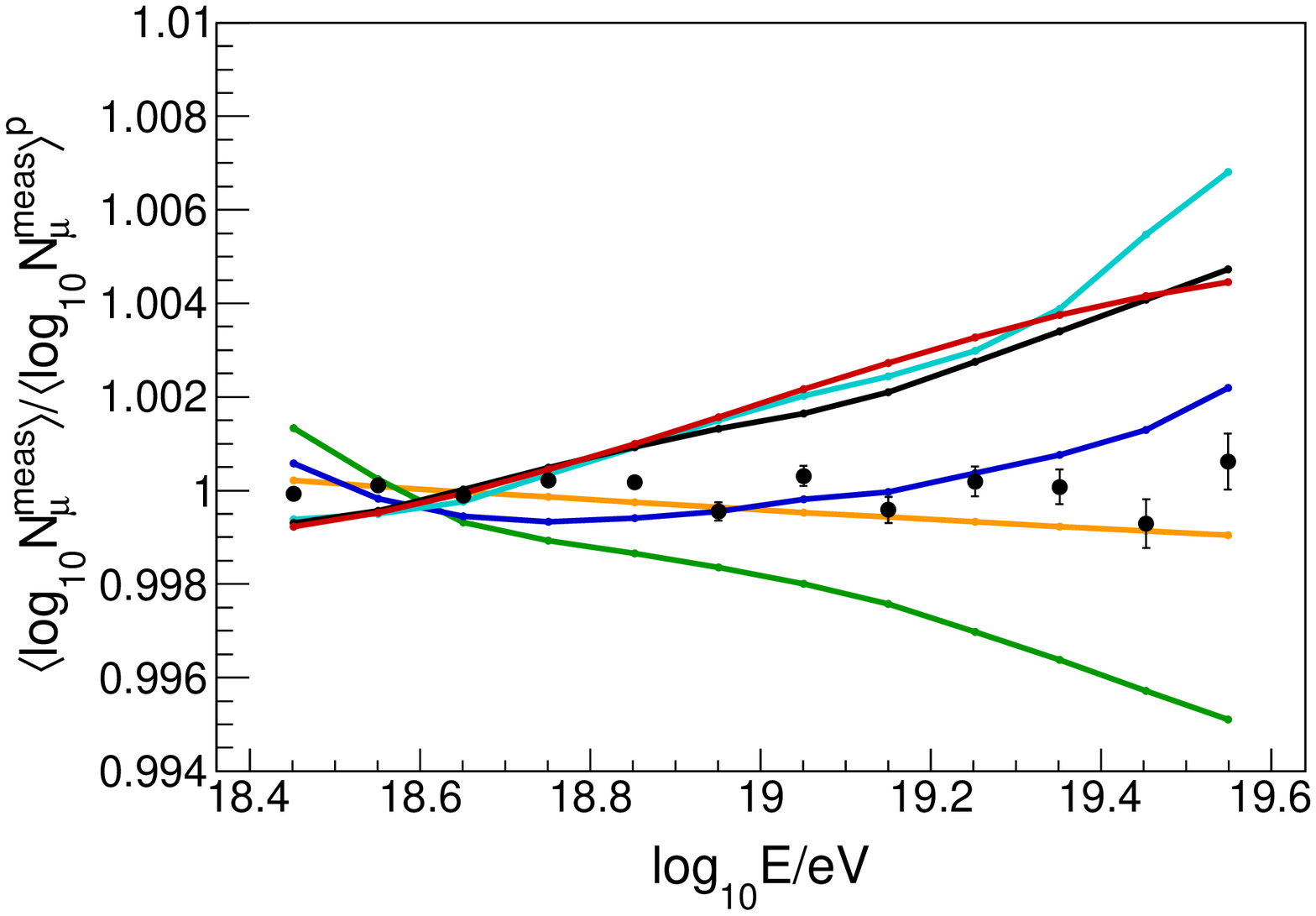}
  }
  \subfloat[]{
    \includegraphics[width=0.47\textwidth]{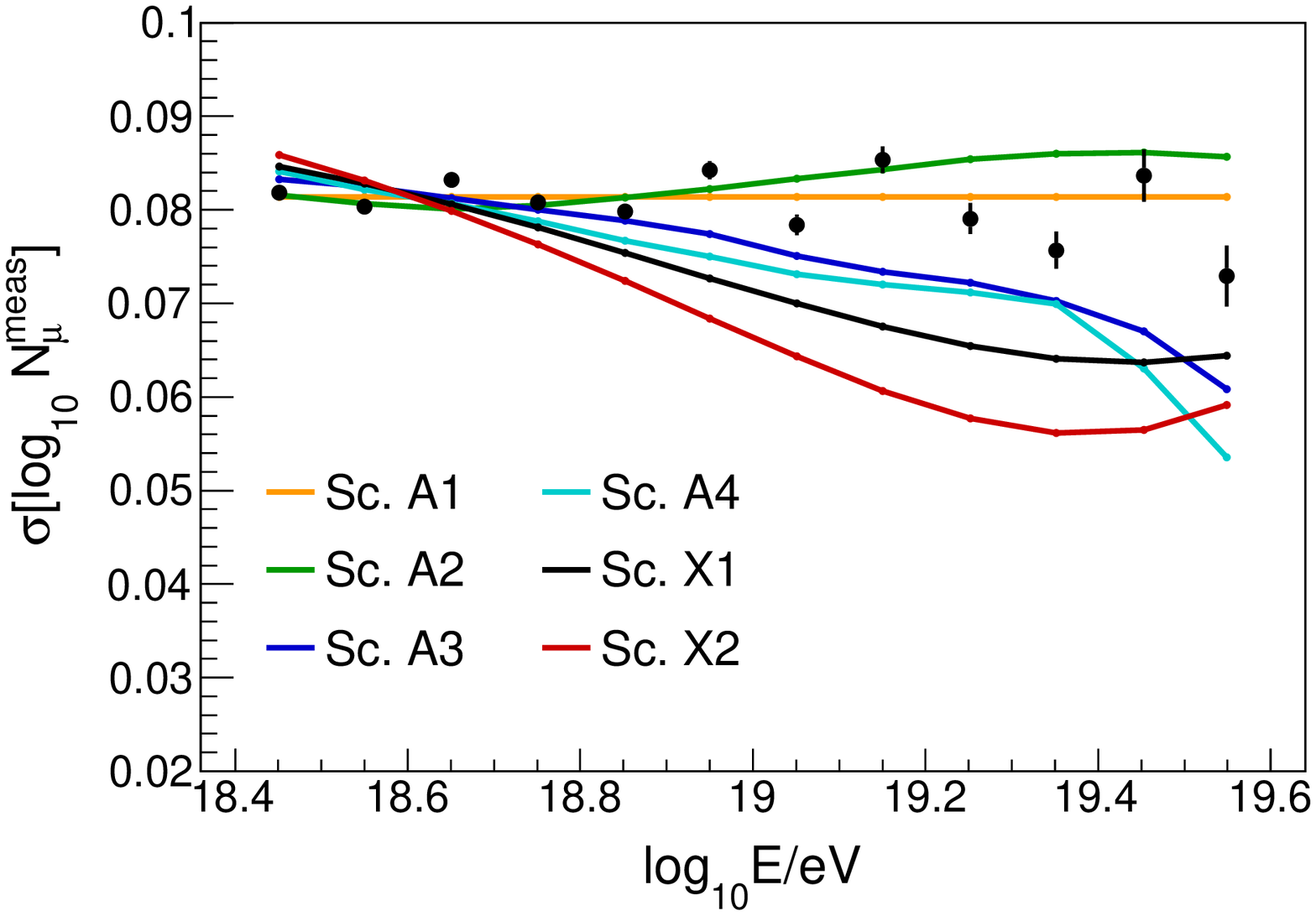}
  }

  \caption{Black dots show the simulated (a) \lgnmumeanexp and (b) \lgnmurmsexp using scenario A1 as the true one. The colored lines show the results of the fit for each one of the test scenarios. The values of \lgnmumeanexp are divided by the corresponding value of pure proton composition for better visualization.}
  \label{fig:fit:example}
\end{figure*}

In next step, we perform a \cchi fit using the model described by~\cref{eq:model:mean:tot,eq:model:rms:sum}, with $a$ and $b$ as free parameters of the fit, for all the composition scenarios. The scenarios which are not the true one are named \textit{test scenarios}. \cref{fig:fit:example} shows one realization of these fits in which scenario A1 was used as the \textit{true scenario} to generate the black dots. The fit of the \lgnmumeanexp with energy (\cref{fig:fit:example}) sets the best value of $a$ and the minimal value of $\chi^2(a)$. The fit of the \lgnmurmsexp with energy (\cref{fig:fit:example}) sets the best value of $b$ and the minimal value of $\chi^2(b)$. In all fits, $D_{E} = 0.920$, $D_{A} = 0.0354$ and $ \sigma [\log_{10} N_{\mu} ]_{\rm Fe} = 0.0265$. Each line in~\cref{fig:fit:example} is the fit of one out of the six composition scenarios. We compared all scenarios (\textit{test scenarios}) to the \textit{true scenario}.

\begin{figure*}
  \subfloat[]{
    \includegraphics[width=0.47\textwidth]{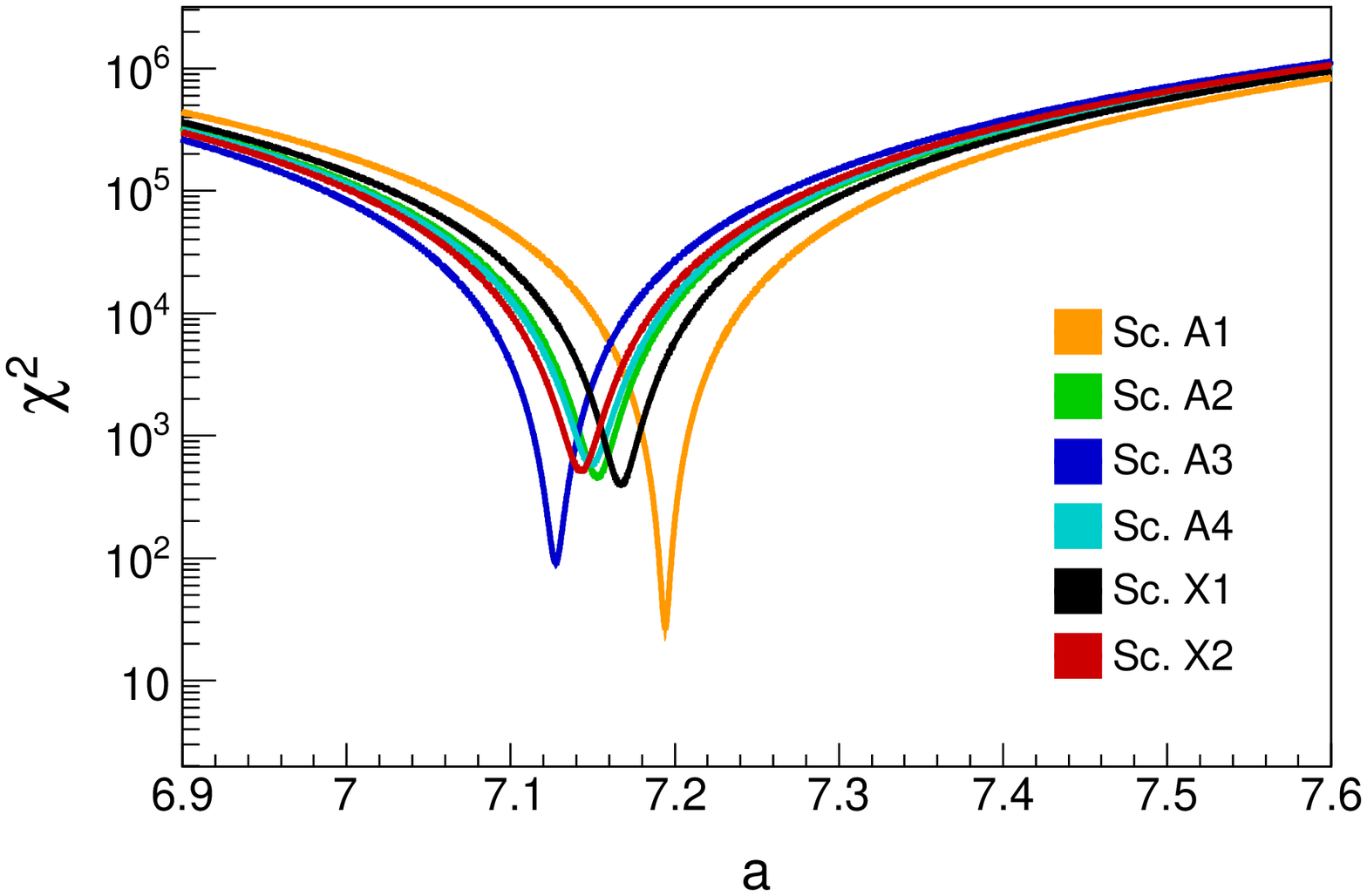}
  }
  \subfloat[]{
    \includegraphics[width=0.47\textwidth]{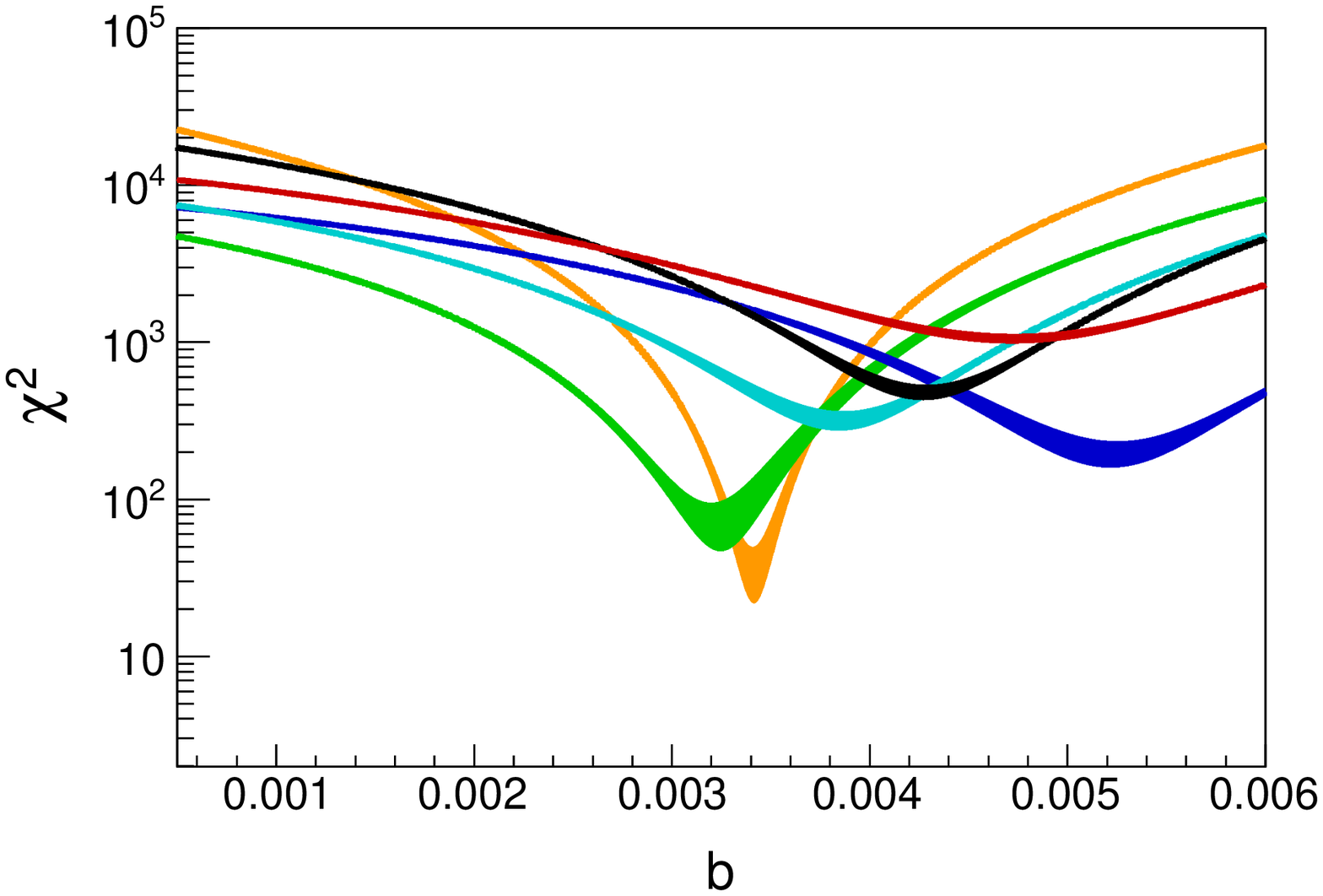}
  }

  \caption{\cchi as a function of the parameters $a$ (a) and $b$ (b) for all the scenarios. The scenario A1 is the \textit{true scenario}. The colored bands represent one standard deviation around the mean for a set of 500 realizations.}
  \label{fig:fit:chi}
\end{figure*}

\begin{figure*}
  \subfloat[Scenario A1]{
    \includegraphics[width=0.47\textwidth]{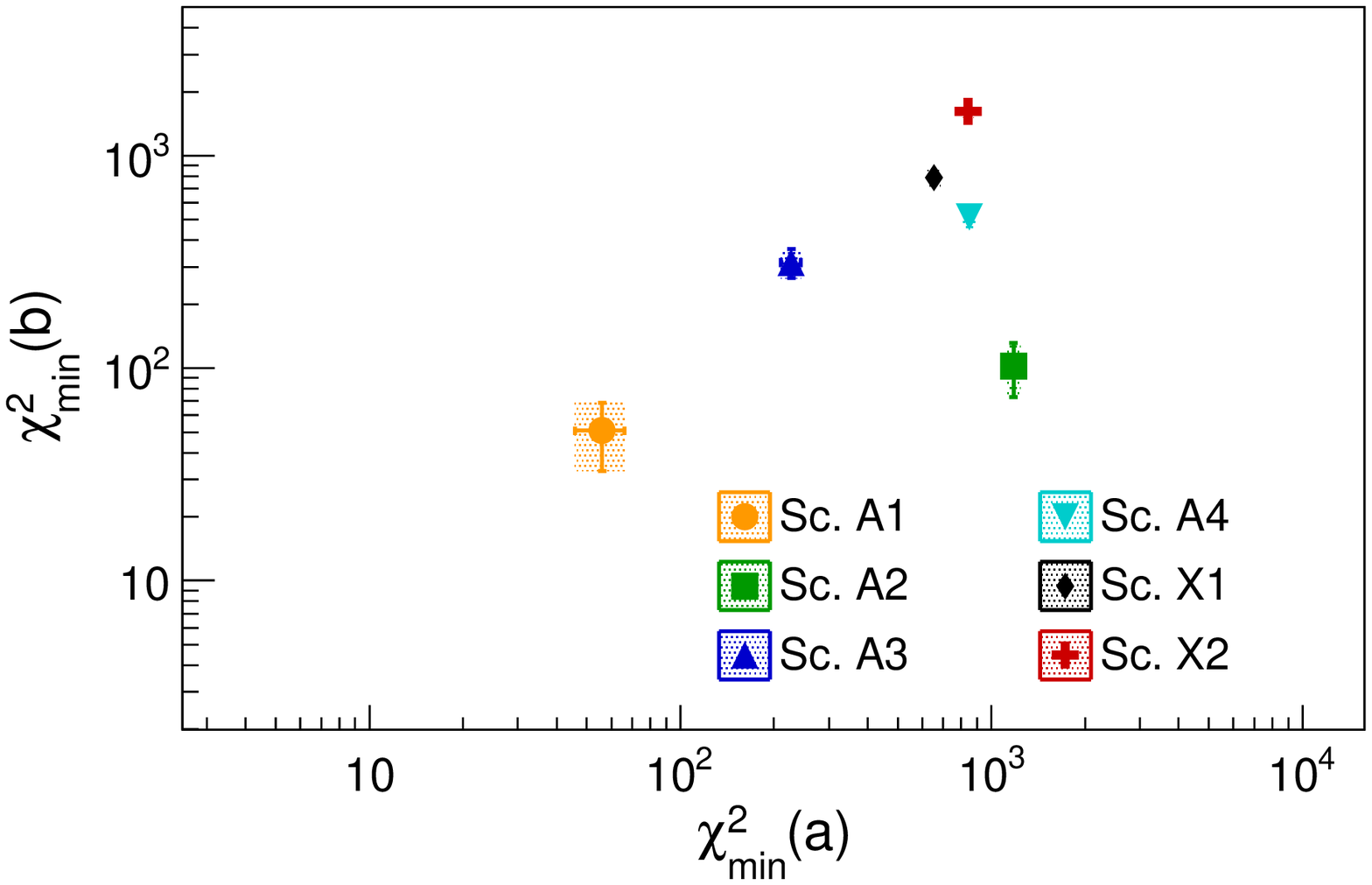}
  }
  \subfloat[Scenario A2]{
    \includegraphics[width=0.47\textwidth]{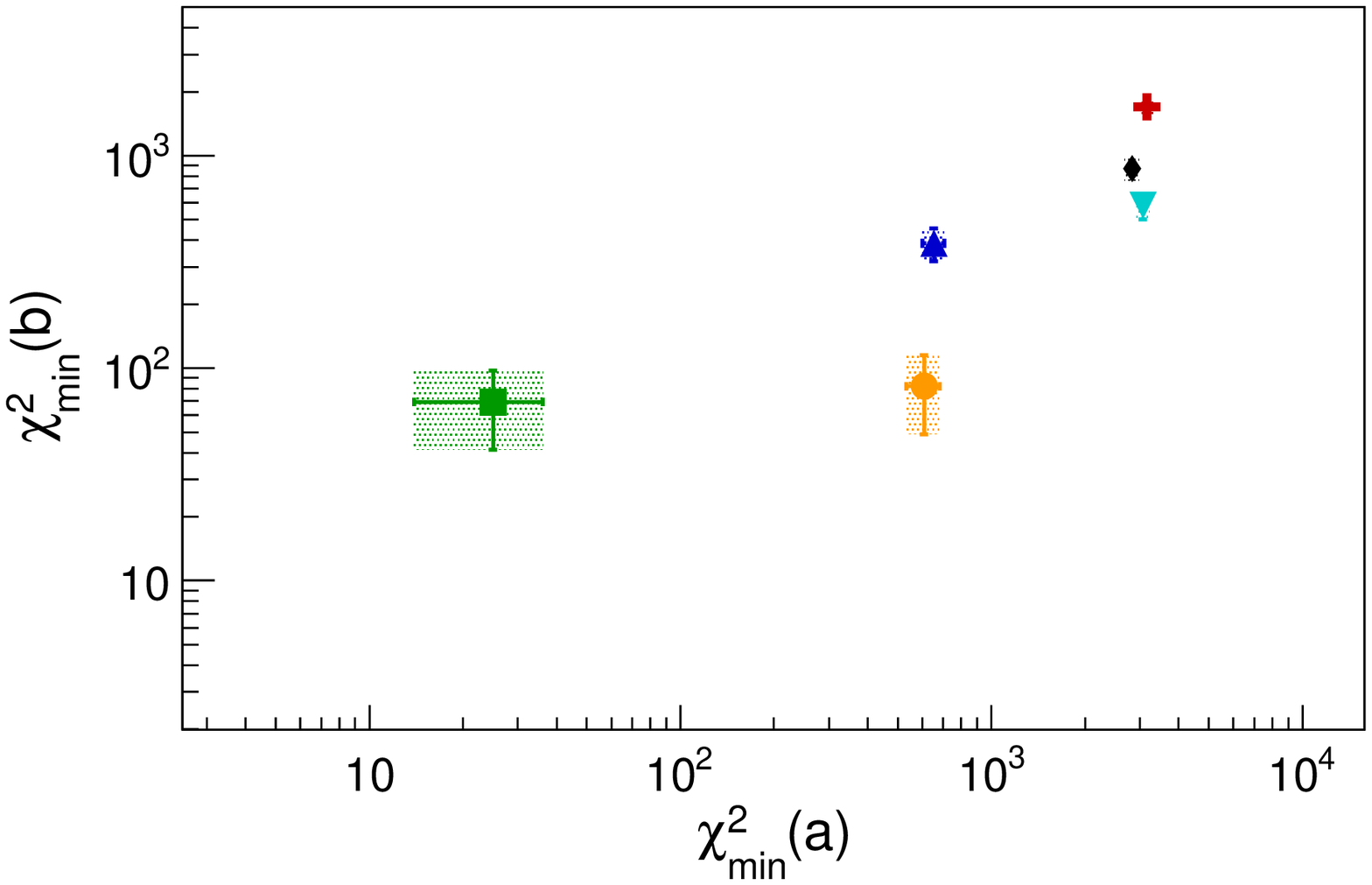}
  }\\
  \subfloat[Scenario A3]{
    \includegraphics[width=0.47\textwidth]{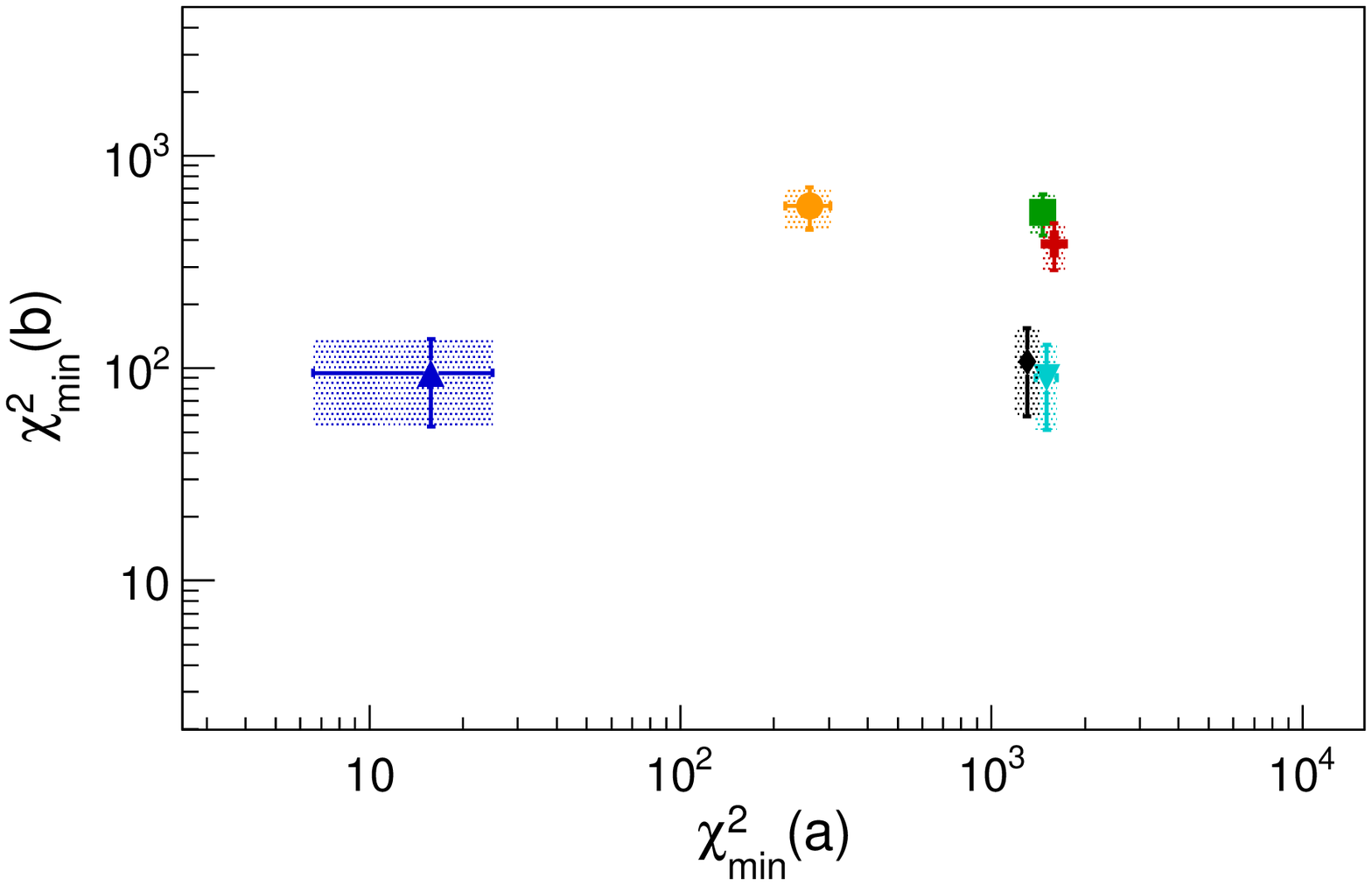}
  }
  \subfloat[Scenario A4]{
    \includegraphics[width=0.47\textwidth]{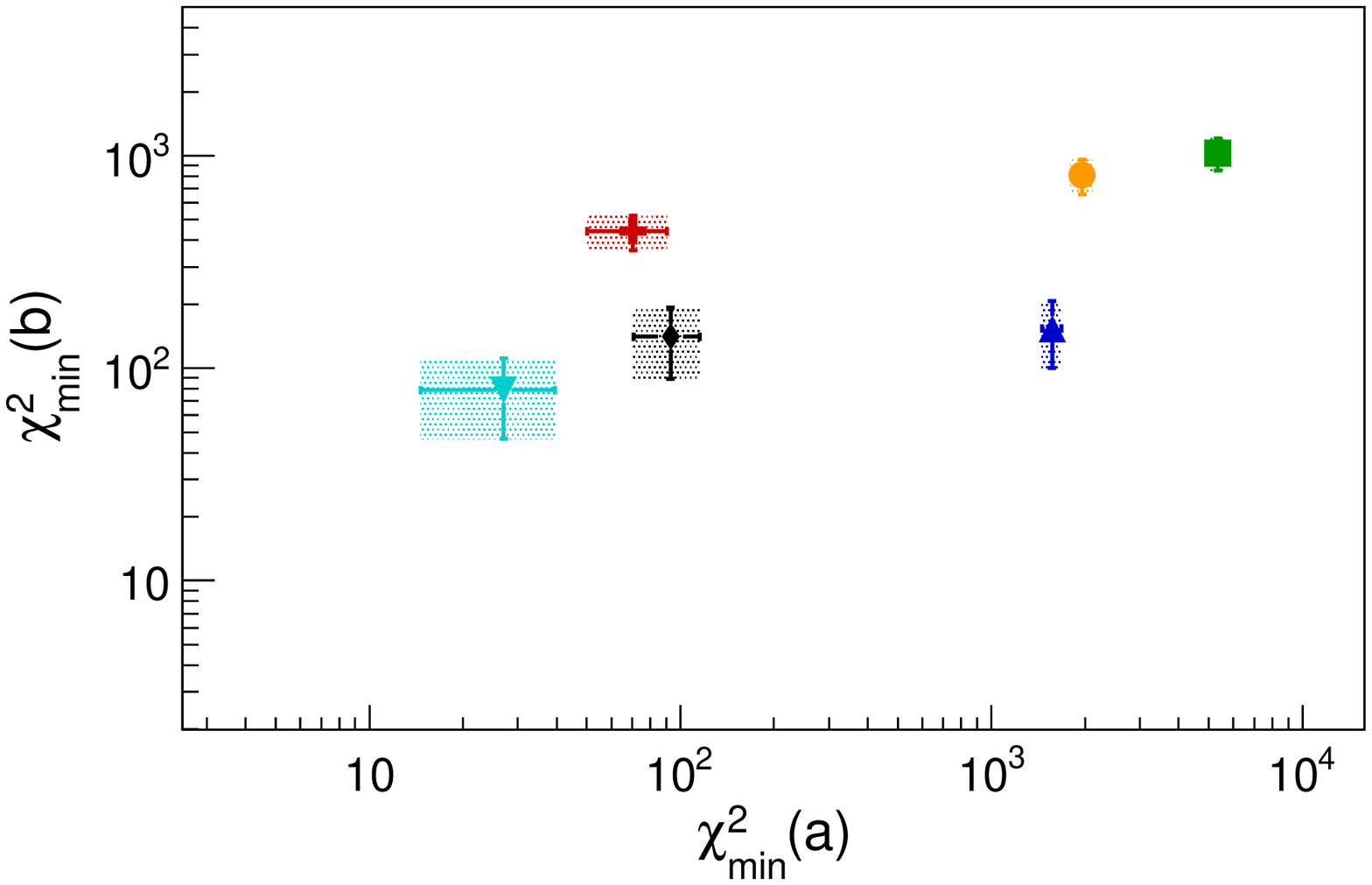}
  }\\
  \subfloat[Scenario X1]{
    \includegraphics[width=0.47\textwidth]{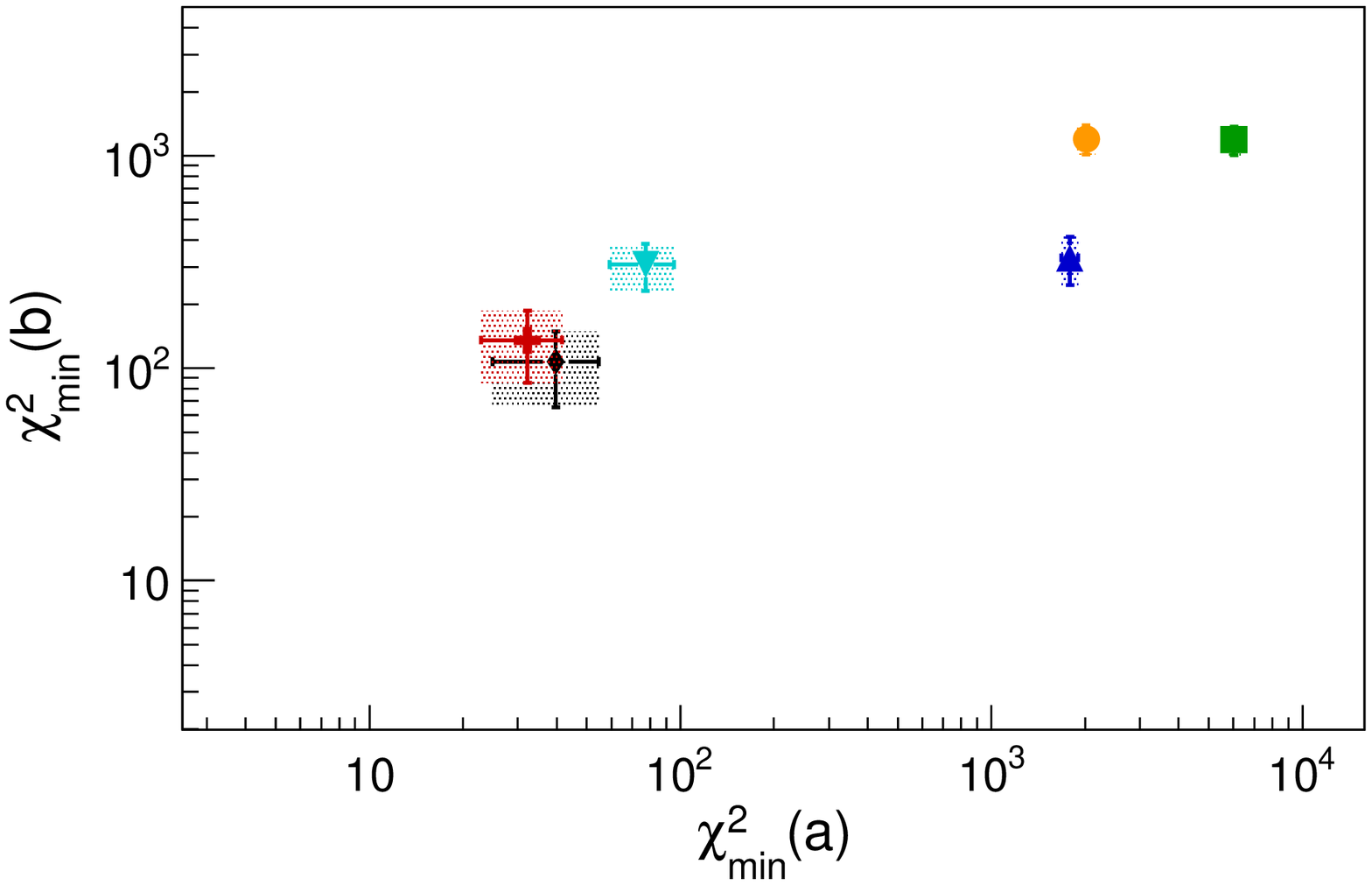}
    \label{fig:min:xmax:a}
  }
  \subfloat[Scenario X2]{
    \includegraphics[width=0.47\textwidth]{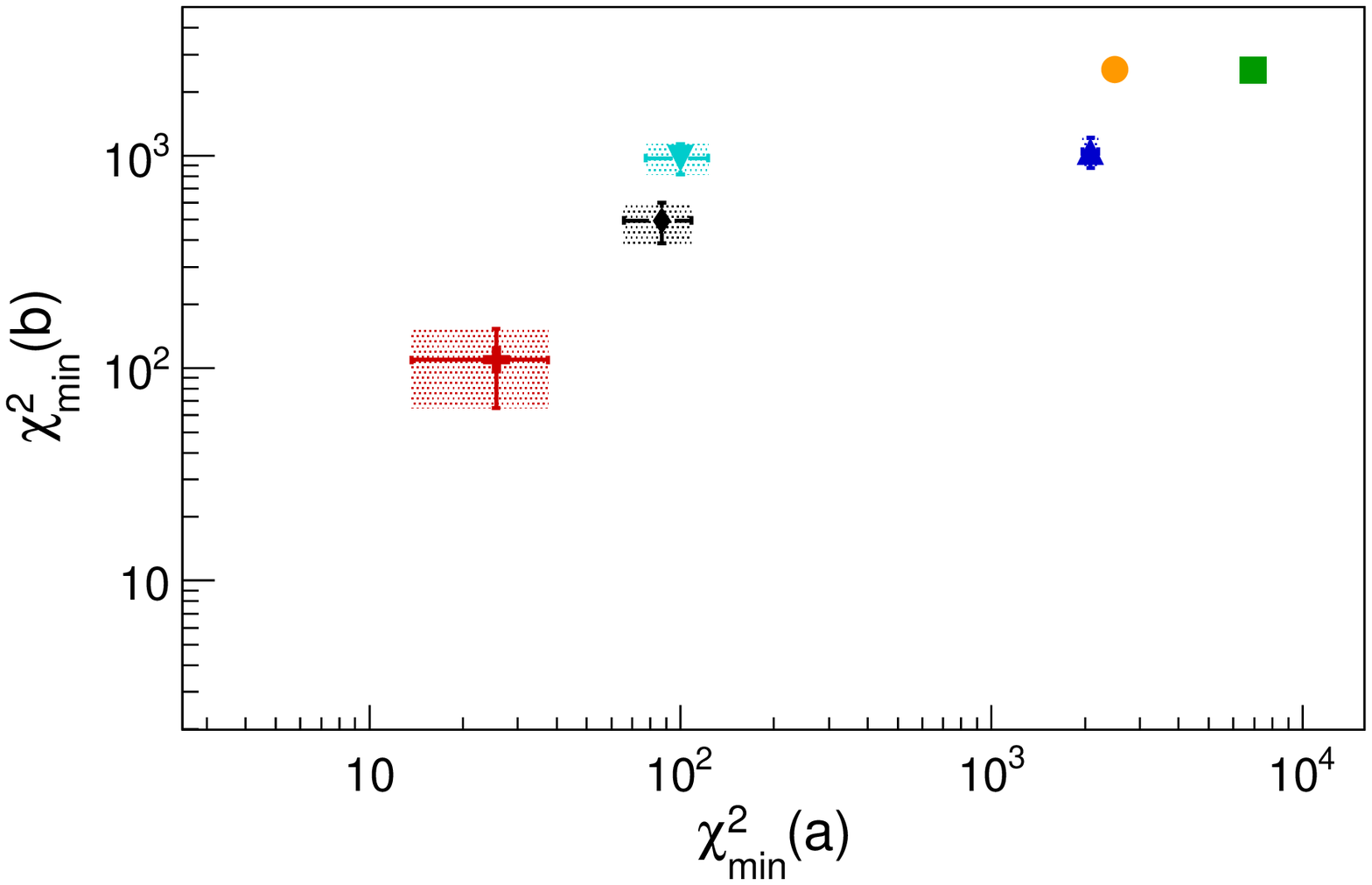}
    \label{fig:min:xmax:b}
  }

  \caption{\chiminalpha \textit{vs} \chiminbeta for all composition scenarios.}
  \label{fig:min}
\end{figure*}

A simple $\chi^2$ comparison finds the \textit{test scenarios} which best fit the data generated with the \textit{true scenario}. The average value of $\chi^2$ as a function of the fitted parameters $a$ and $b$ is shown in~\cref{fig:fit:chi} for a set of 500 realizations. In this case, it is clear that the scenario A1 better describes the \lgnmumeanexp and \lgnmurmsexp evolution with energy because of the smaller values of $\chi^2_{\rm min}(a)$ and $\chi^2_{\rm min}(b)$. \cref{fig:min} shows the plots of $\chi^2_{\rm min}(a)$ \textit{versus} $\chi^2_{\rm min}(b)$ for all six scenarios as the \textit{true scenarios}. The error bars represent one standard deviation around the mean for 500 realizations. All scenarios, except in X1 case, can be discriminated by the smallest  $\chi^2_{\rm min}(a)$ and $\chi^2_{\rm min}(b)$. In other words, the  \textit{true scenario} is the one with the $\chi^2_{\rm min}(a)$-$\chi^2_{\rm min}(b)$ point closer to the left-down corner. Note that only  $\chi^2_{\rm min}(a)$ or only $\chi^2_{\rm min}(b)$ cannot alone discriminate most of the scenarios. In the case of X1 as \textit{true scenario} one can see that, even if it is not possible to discriminate scenario X1 and X2, it is still possible to discriminate the \xmax scenarios from the astrophysical ones.

%=====================================================================
% fit example of log nmu versus energy and rms nmu versus energy

%=====================================================================
% chi2 vs a,b

%=====================================================================
% chi min

\subsection{Sensitivity to the systematic uncertainties on energy scale and absolute number of muons}
\label{sec:results:syst}

As mentioned above, the greatest obstacles in interpreting \nmu data currently are the systematic uncertainties in the theoretical description and reconstruction of air showers. In this section we demonstrate that the procedure proposed in the previous section to discriminate between composition scenarios is stable under systematic changes of absolute \nmu prediction and of energy scale.

\begin{figure*}
  \subfloat[Scenario A1]{
    \includegraphics[width=0.47\textwidth]{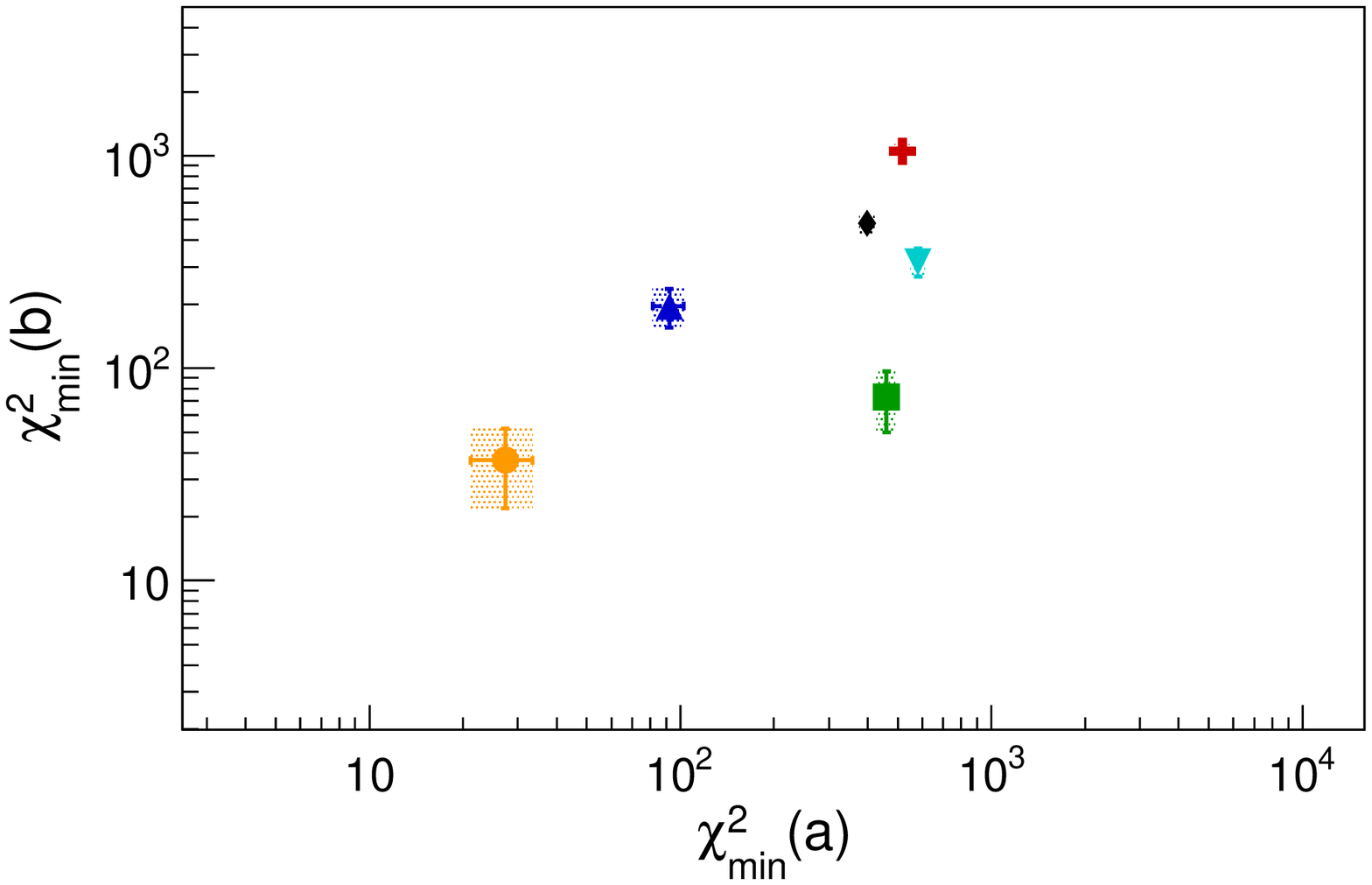}
    \label{fig:min:nmu:a130}
  }
  \subfloat[Scenario A1]{
    \includegraphics[width=0.47\textwidth]{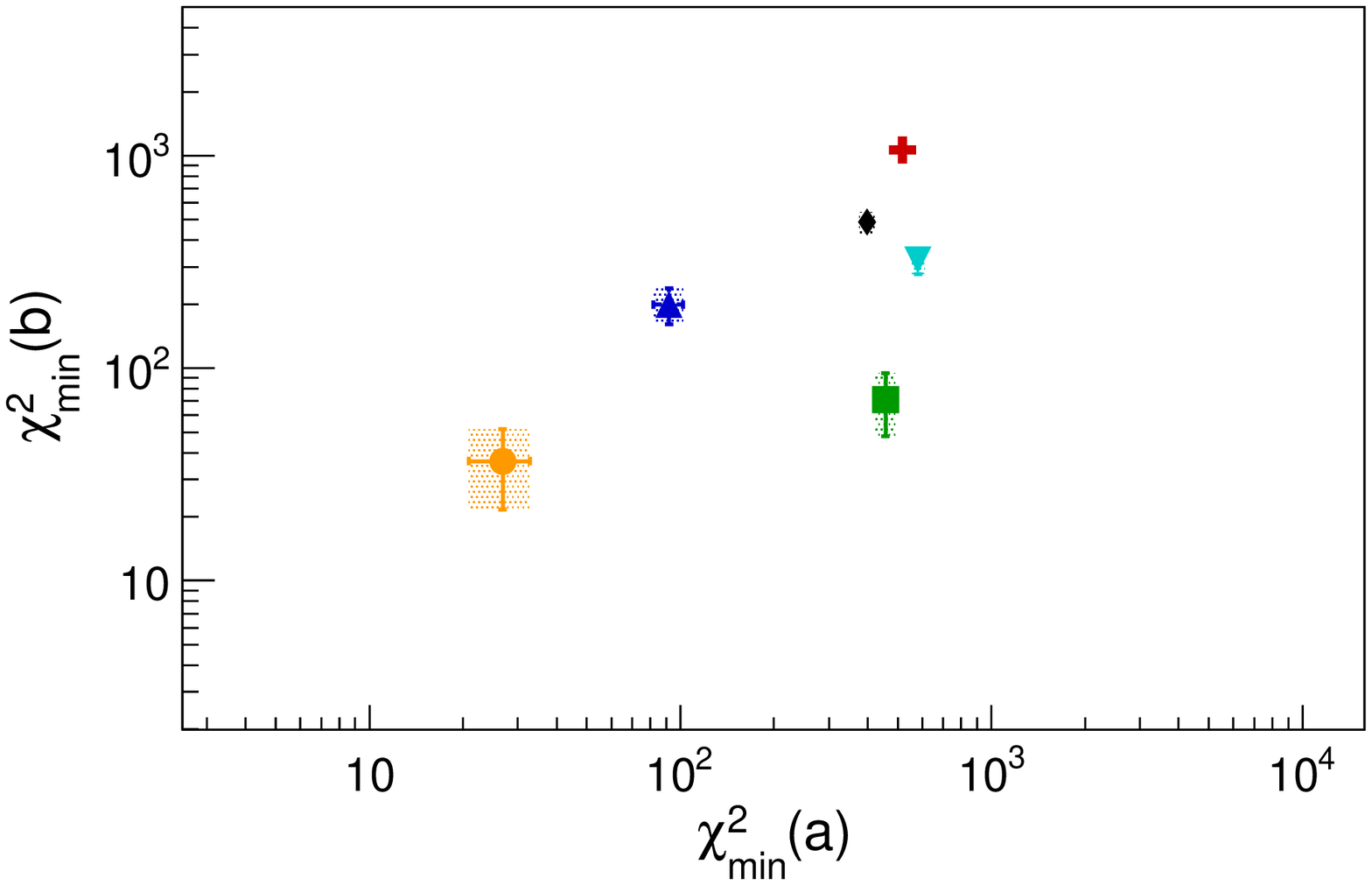}
    \label{fig:min:nmu:a160}
  }

  \subfloat[Scenario X1]{
    \includegraphics[width=0.47\textwidth]{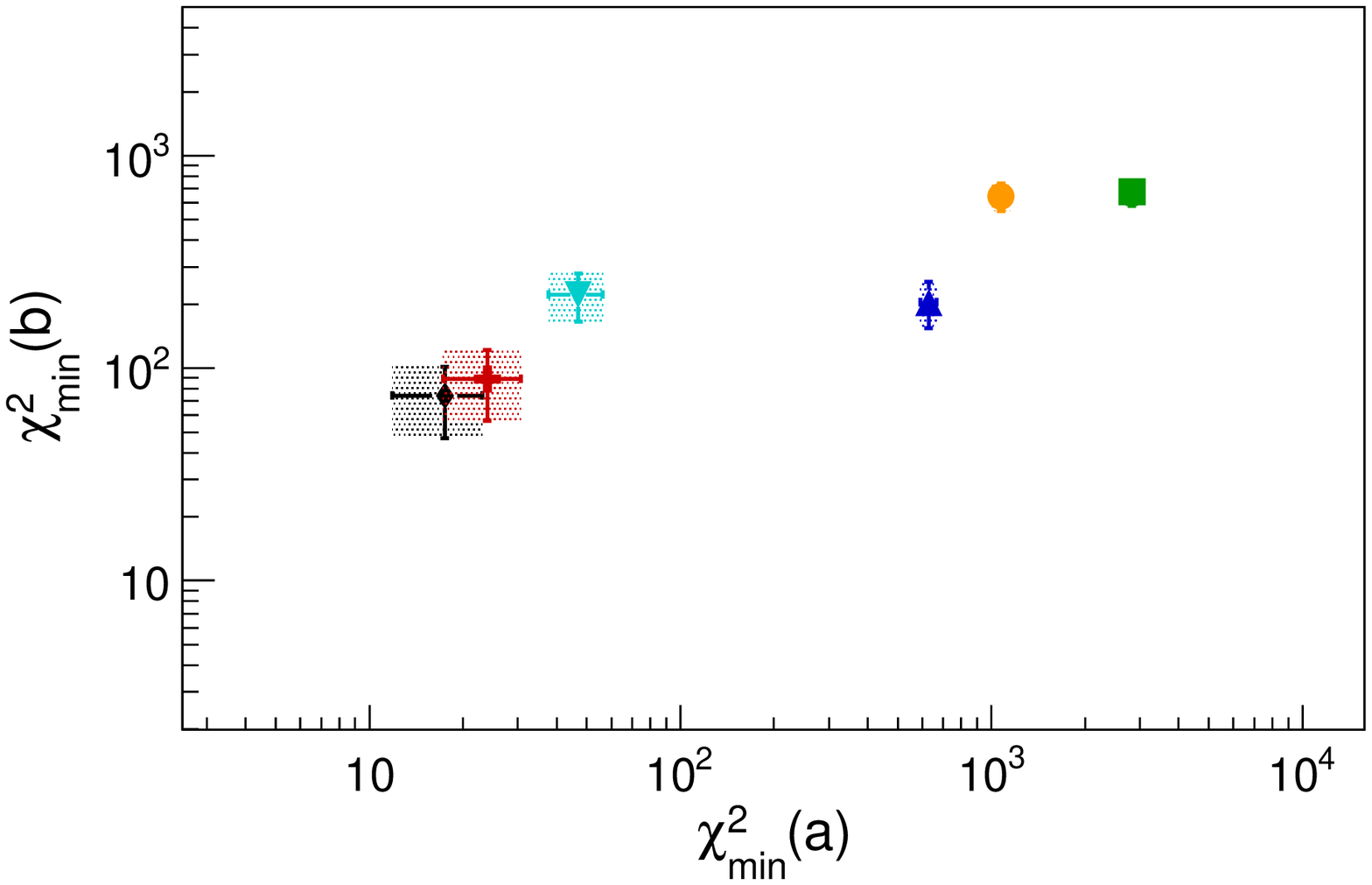}
    \label{fig:min:nmu:x130}
  }
  \subfloat[Scenario X1]{
    \includegraphics[width=0.47\textwidth]{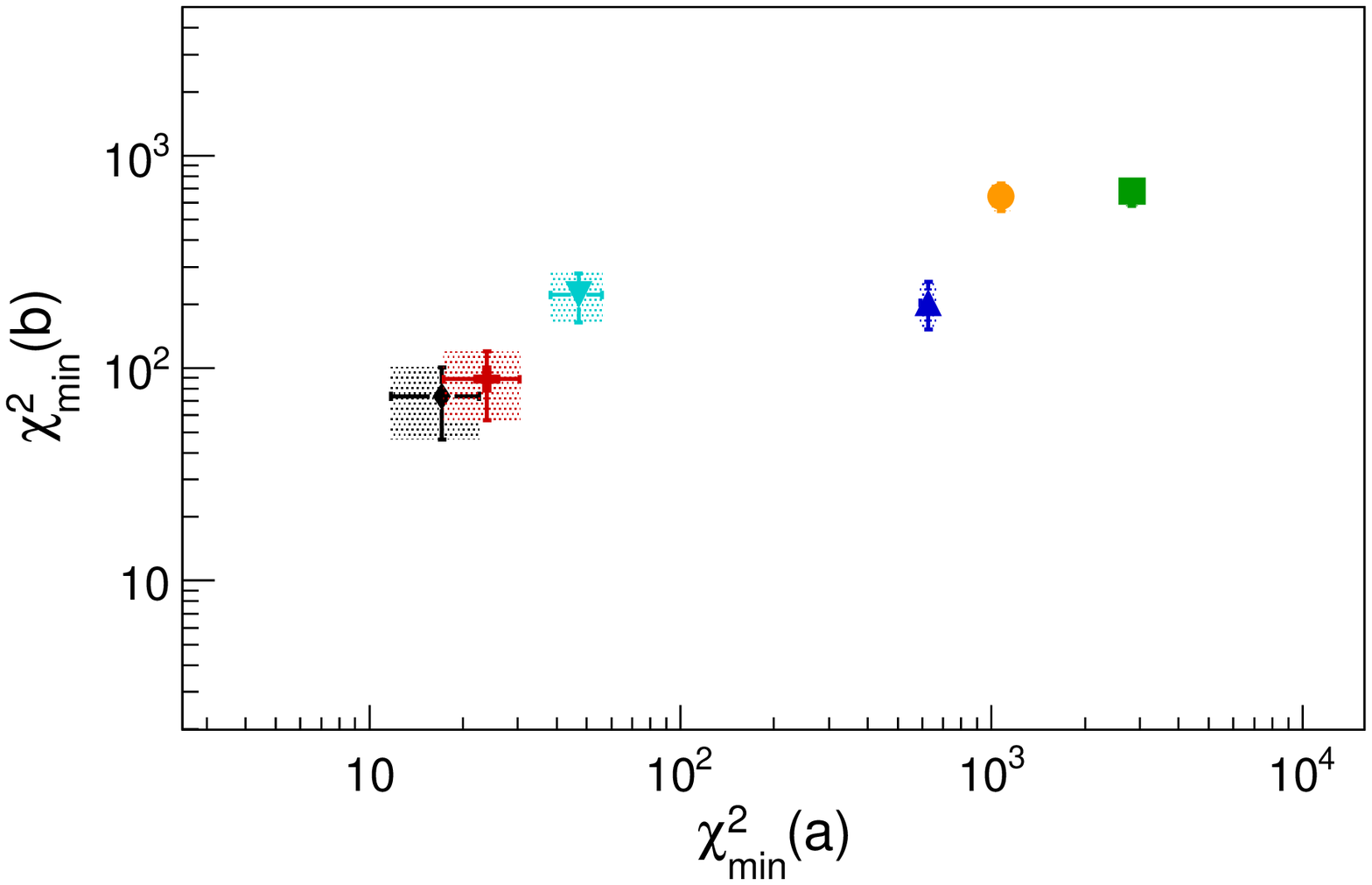}
    \label{fig:min:nmu:x160}
  }

  \caption{\chiminalpha \textit{vs} \chiminbeta for the true scenario A1 with (a) \fnmu$=1.3$ and (b) \fnmu$=1.6$ and for the true scenario X1 with (c) \fnmu$=1.3$ and (d) \fnmu$=1.6$ (see text).}
  \label{fig:min:nmu}
\end{figure*}

The systematic uncertainties in \nmu scale were tested by applying the rescaling factor \fnmu in \nmuexp generated by simulations. The examination of~\cref{eq:model:mean} shows that a rescaling factor on \nmu corresponds to an additive term in $a$. This is how the systematic effect on \nmuexp is incorporated in the simple description of \lgnmumeanexp. \cref{eq:model:rms:lna} shows that $b$ does not depend on \fnmu. If only \lgnmumeanexp changes by an additive term when \fnmu is applied, it is clear that \chiminalpha and \chiminbeta are independent of \fnmu. \cref{fig:min:nmu} shows the values of \chiminalpha and \chiminbeta for true scenarios A1 and X1 and \fnmu$=1.3$ and $1.6$, as examples. One can see that the values of \chiminalpha and \chiminbeta are stable under systematic changes in \nmuexp.

\begin{figure*}
  \subfloat[]{
    \includegraphics[width=0.47\textwidth]{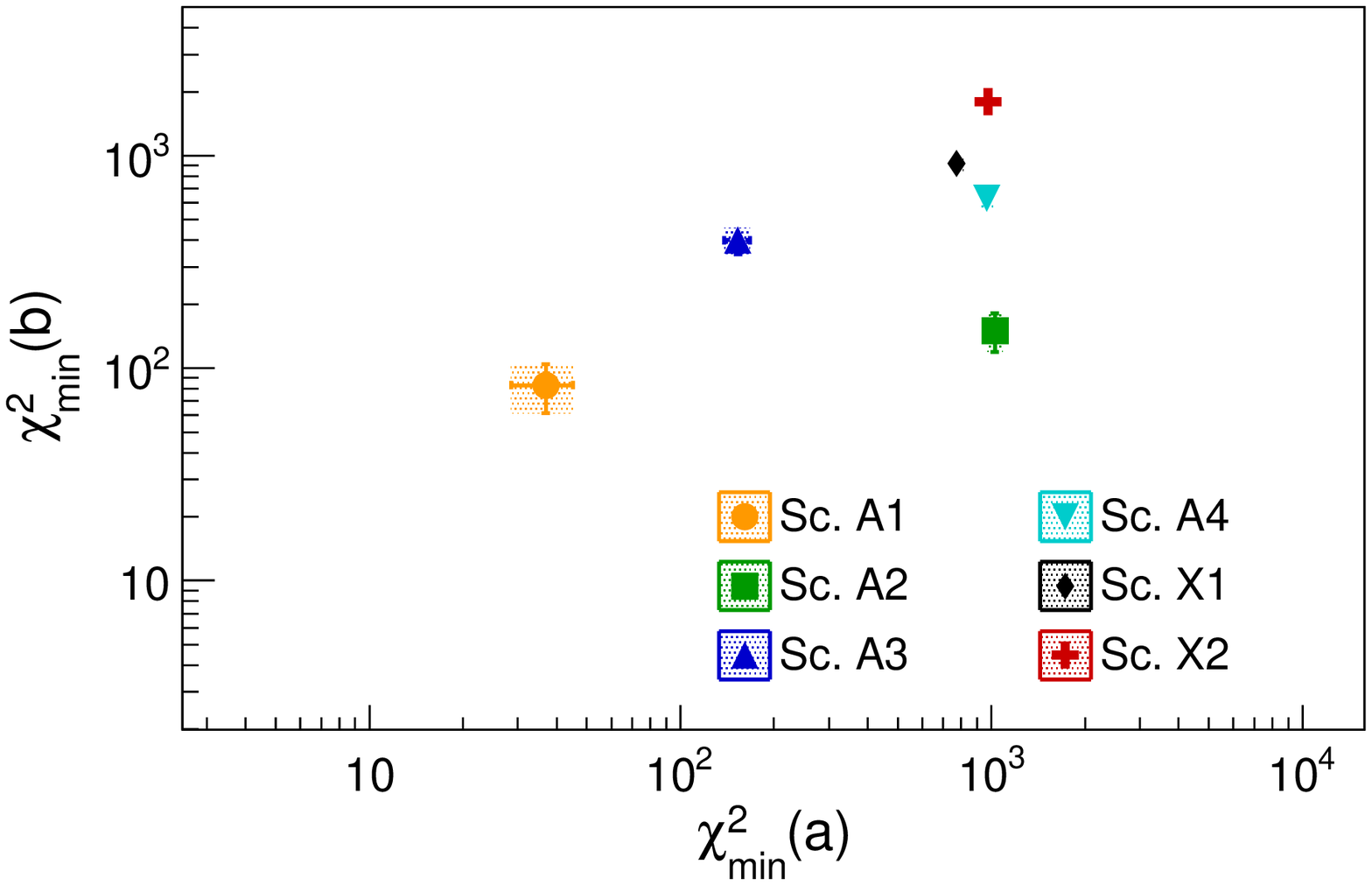}
    \label{fig:min:en:a085}
  }
  \subfloat[]{
    \includegraphics[width=0.47\textwidth]{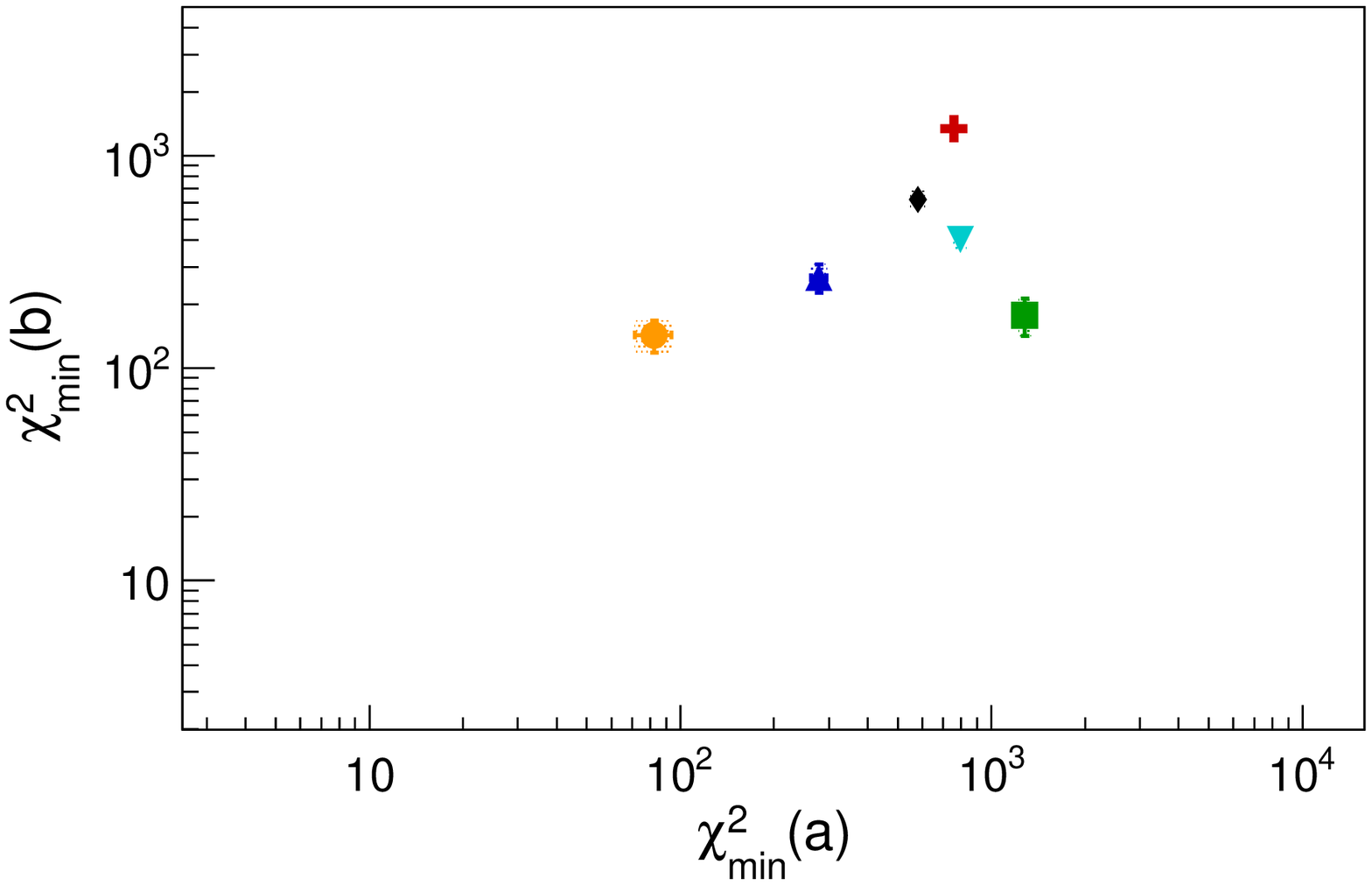}
    \label{fig:min:en:a115}
  }

  \subfloat[]{
    \includegraphics[width=0.47\textwidth]{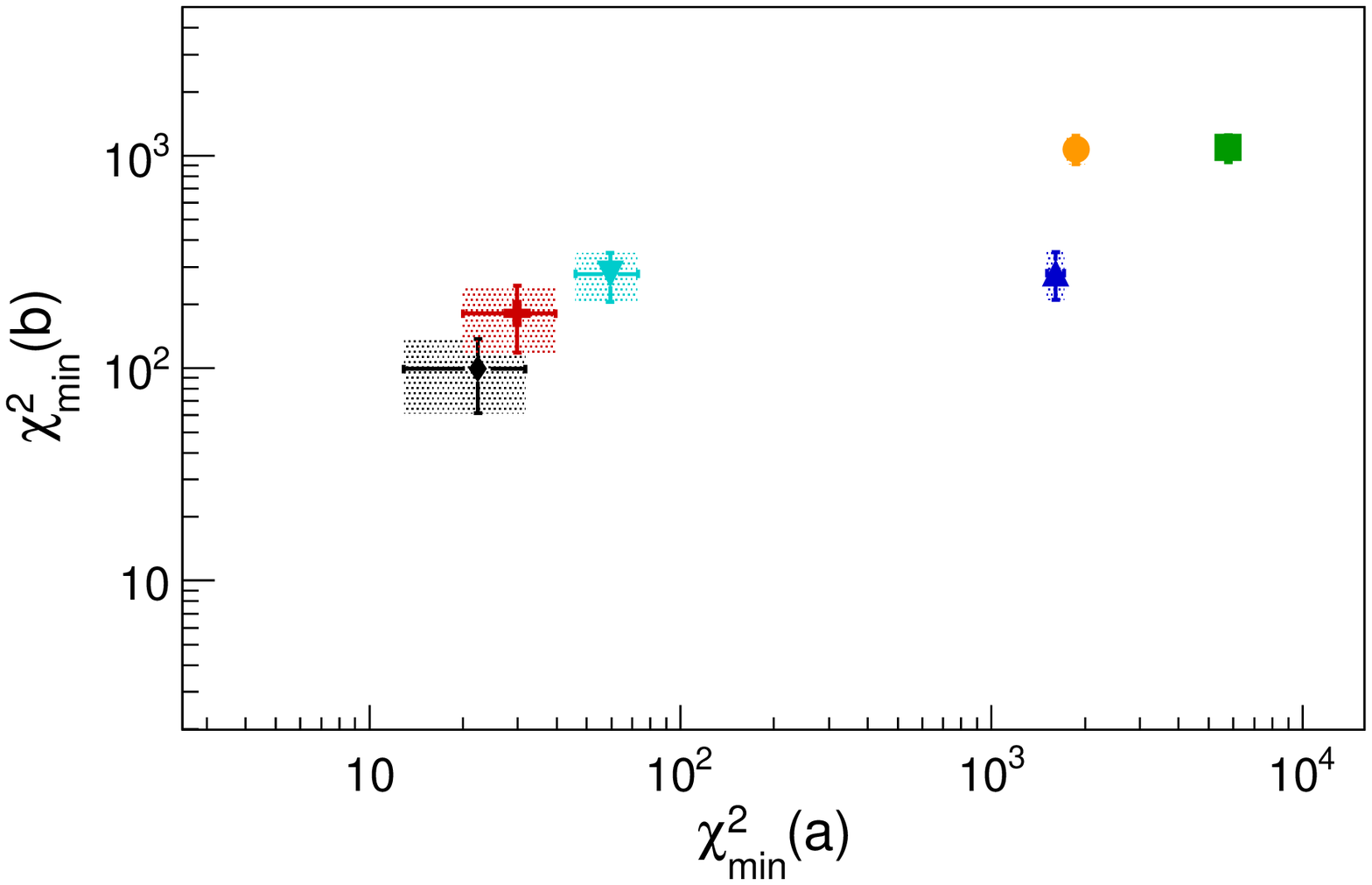}
    \label{fig:min:en:x085}
  }
  \subfloat[]{
    \includegraphics[width=0.47\textwidth]{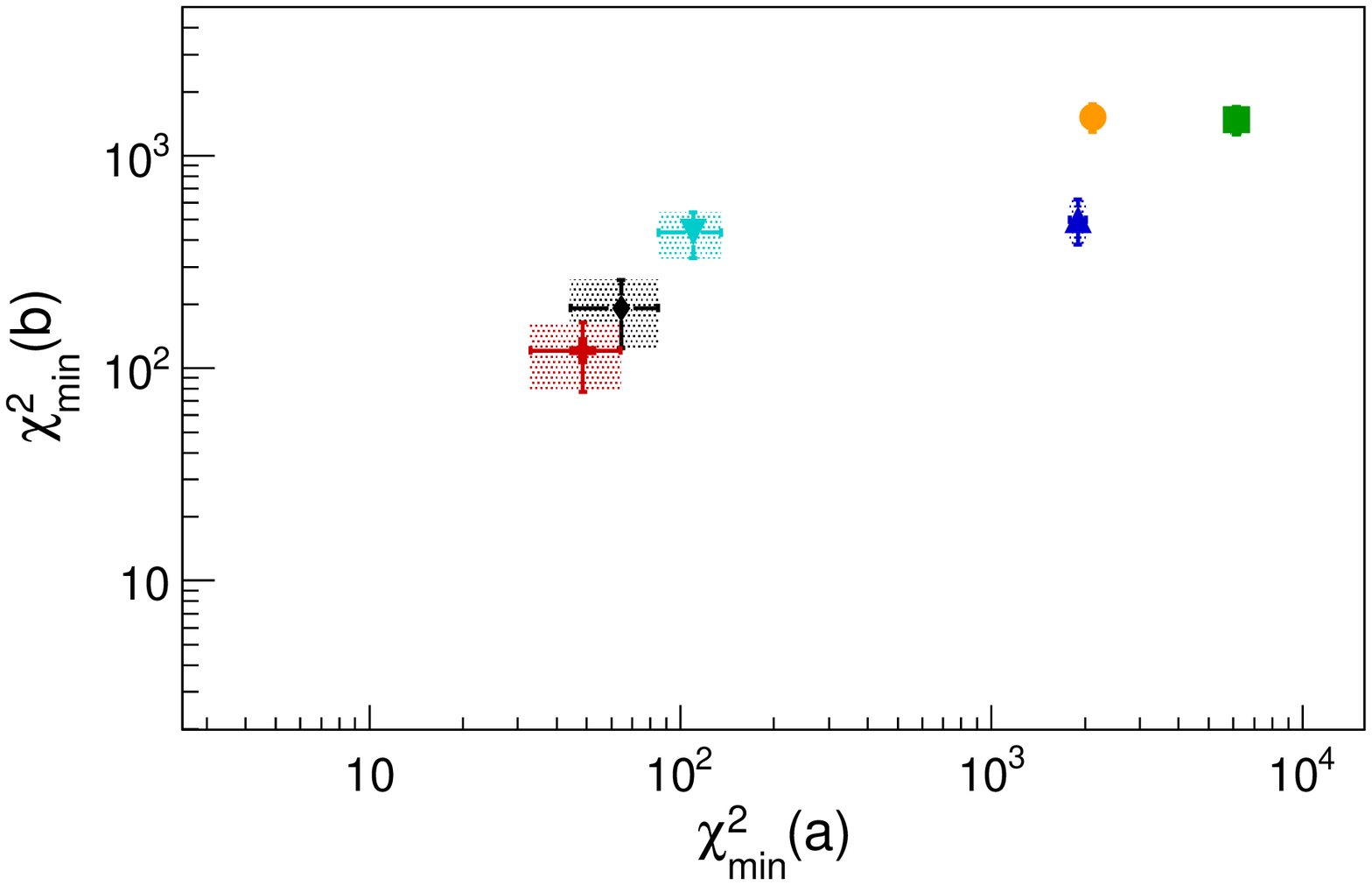}
    \label{fig:min:en:x115}
  }

  \caption{\chiminalpha \textit{vs} \chiminbeta for the true scenario A1 with (a) \fen$=0.85$ and (b) \fen$=1.15$ and for the true scenario X1 with (c) \fen$=0.85$ and (d) \fen$=1.15$ (see text).}
  \label{fig:min:en}
\end{figure*}

The energy scale effect was tested by including a rescaling factor \fen in simulated energy of each shower. The same analysis of~\cref{eq:model:mean,eq:model:rms:lna} reveals that $a$ and $b$ accommodate the systematic effects in energy as additive terms. All additive terms are canceled in the $\chi^2$ comparison resulting in the independence of the conclusions under systematic effects. In \cref{fig:min:en} we show the \chiminalpha and \chiminbeta for true scenarios A1 and X1 and for \fen$=0.85$ and $1.15$, which represents a systematic uncertainties of $15$\% in energy. Again, it can be observed that the values of \chiminalpha and \chiminbeta would not lead to a different conclusion, and therefore, the results of the method would be stable under energy shifts.

%=====================================================================
%systematics study

%%%%%%%%%===================================%%%%%%%%%%%%%%%

\section{Conclusion}
\label{sec:conclusion}

In this paper, we have analyzed the muon content of air showers, proposed a parametrization of the first two moments of the number of muons with energy and primary particle mass and showed how the measured mean and $\sigma$ of \lgnmu can be used to discriminate between composition scenarios.

We proposed a model to describe \lgnmumeanexp and \lgnmurmsexp as a function of energy and primary particle mass ($A$). This model was conceived to keep the most relevant hadronic interaction uncertainties concentrated in only two parameters ($a$ and $b$). We have validated the model with Monte Carlo simulation of the air shower and its capability to describe the \lgnmuexp moments was proven.

Six composition scenarios were considered. The particle flux predicted by these scenarios was transformed into the corresponding \lgnmumeanexp and \lgnmurmsexp evolution with energy. The \lgnmumeanexp and \lgnmurmsexp evolution with energy was fitted using the proposed parametrization. A comparison of the \lgnmumeanexp and \lgnmurmsexp model using a simple $\chi^2$ test allows the discrimination between the scenarios. The discrimination is effective even considering the systematic uncertainties on the \nmu prediction and on energy scale uncertainty.

The effect of the systematic in the \nmu number and energy reconstruction was studied for constant values of the uncertainty with energy. This choice is justified by the narrow energy interval used in the analysis. Abrupt changes of the systematic uncertainties with energy could change the conclusion drawn here since \lgnmuexp is assumed to be fixed in the proposed model.% Detector resolution and acceptance were also neglected in our analysis. However, a good knowledge of the detector response would allow the unfolding or unbiasing of the data as done for the \xmax analysis~\cite{Aab2014}.

The upgrade of Telescope Array~\cite{ta:extension} and Pierre Auger Observatory~\cite{Engel2015}, to be constructed in the next few years, will for the first time allow precise measurements of the muon component of air showers for energies above \energy{18} eV. This will open up a new window of analyses and tests in astroparticle physics. Once data is acquired, the parametrization proposed here could be tested and if proven to be right, the analysis method proposed in~\cref{sec:analysis} could be used to find the most probable composition scenario in the energy range from $10^{18.4}$ to $10^{19.6}$ eV.

%==============================================
\section*{Acknowledgements}
We thank St\'ephane Coutu and Carola Dobrigkeit for the review of the manuscript on behalf of the Pierre Auger Collaboration. RRP thanks the financial support given by FAPESP (2014/10460-1). RC and MP gratefully acknowledge the financial support by Funda\c{c}\~{a}o para a Ci\^{e}ncia e Tecnologia (SFRH/BPD/73270/2010) and OE, FCT-Portugal, CERN/FIS-NUC/0038/2015. VdS thanks the support of the Brazilian population via CNPq and FAPESP (2010/07359-6, 2014/19946-4).

%=========================================================================================================

\appendix
\section{Parametrization of $R(E,X_{\rm max})$}
\label{sec:app}

As mentioned in~\cref{sec:sim}, the parametrization of $R(E,X_{\rm max})$ can be done by means of full air shower simulations. In this paper, we choose CORSIKA \cite{Heck1998a} as the full Monte Carlo code. First, a set of 200 CORSIKA (version 7.4000) showers for each primary (proton, helium, nitrogen and iron), fixed energy ($E =  10^{18.5}, 10^{19.0}, 10^{19.5}$ eV) and high energy hadronic interaction model (EPOS-LHC \cite{Werner2007} and QGSJetII-04 \cite{Ostapchenko2010}) were generated. The low energy hadronic interaction model is Fluka \cite{Fluka2007} in all cases. Furthermore, the zenith angle was set fixed to $38^{\circ}$ for all showers. This is clearly a simplification, and then it should be stressed that, in a more realistic analysis, the \nmu zenith angle dependence must be treated.

\begin{figure}[]
  \centering
  \includegraphics[width=0.47\textwidth]{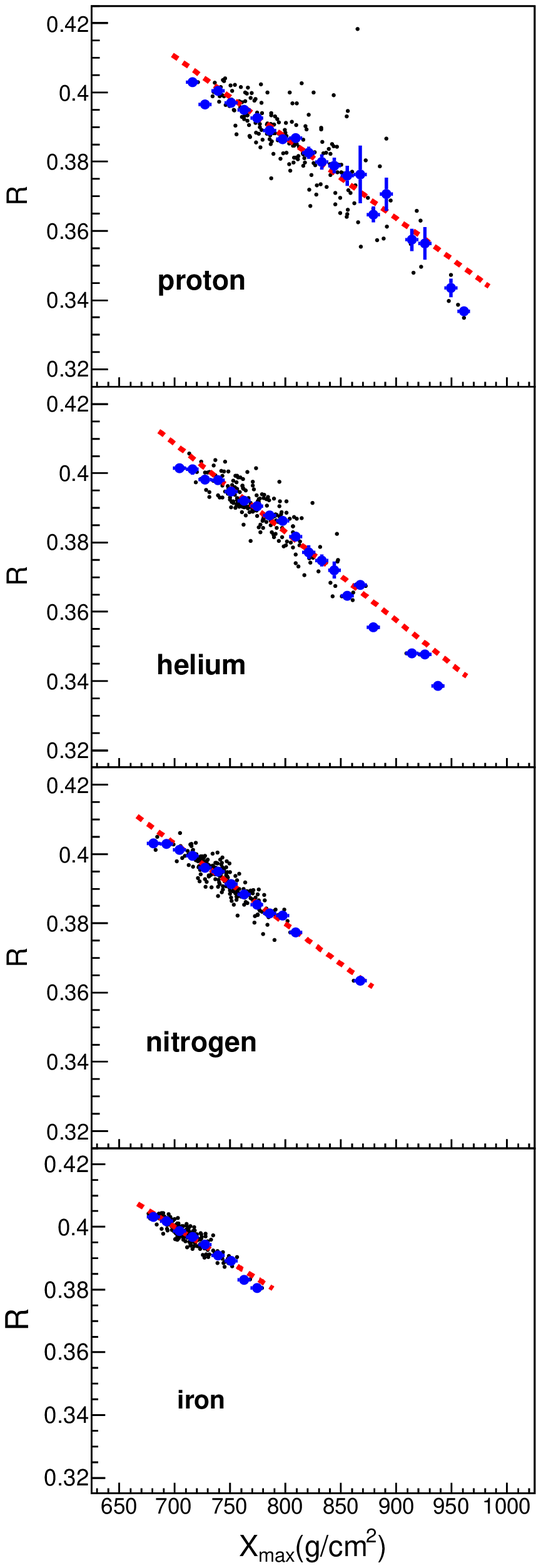}
  \caption{Conversion factor $R$ as a function of \xmax for $E=10^{19.0}$ eV showers with EPOS-LHC as hadronic interaction model. The black dots are individual showers, the blue circles are the profile and dotted red line is the linear fit.}
  \label{fig:toy:conversion:eposlhc}
\end{figure}

In~\cref{fig:toy:conversion:eposlhc} the factor $R$ is shown as a function of \xmax for all primaries, one primary energy, $10^{19.0}$ eV, and one hadronic interaction model, EPOS-LHC. The observed behavior suggests a linear parametrization of the form,

\begin{equation}
  R(E,X_{\rm max}) = p_1(E) \cdot X_{\rm max}+p_0(E) \; ,
\end{equation}

\noindent
where $p_0$ and $p_1$ have to be also parametrized as a function of energy. The values of $p_0$ and $p_1$ for each primary and energy were determined from the linear fit, shown in~\cref{fig:toy:conversion:eposlhc} by a dotted red line.

\begin{figure}[]
  \centering
  \includegraphics[width=0.47\textwidth]{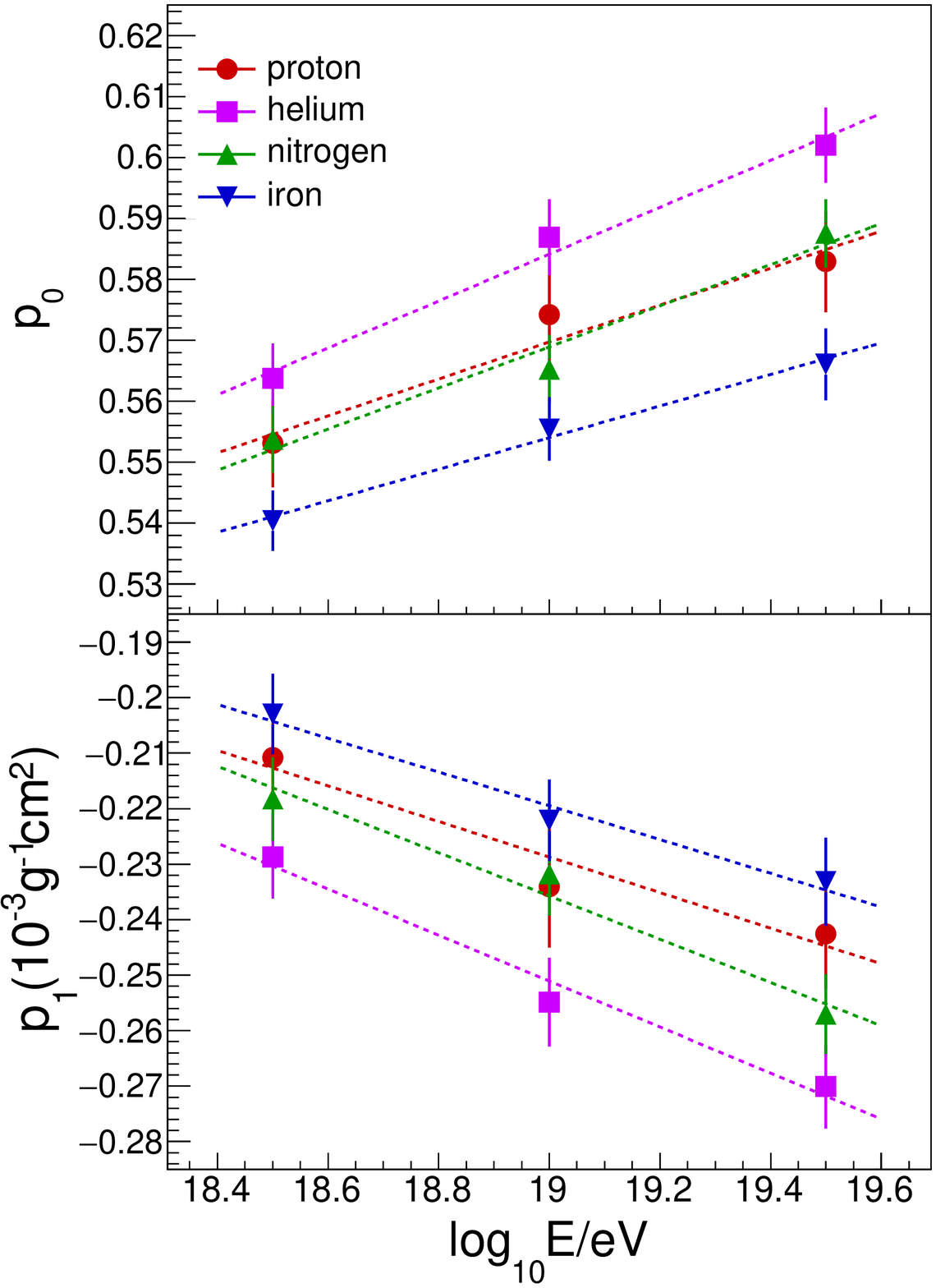}
  \caption{$p_0$ and $p_1$ as a function of \lge (see text) for EPOS-LHC as the hadronic interaction model. The dashed lines are the linear fits represented in~\cref{eq:toy:ps}.}
  \label{fig:toy:param:eposlhc}
\end{figure}

In~\cref{fig:toy:param:eposlhc} $p_0$ and $p_1$ are shown as a function of logarithmic energy for all primaries and the hadronic interaction model EPOS-LHC, as an example. Again, we are able to perform a linear parametrization of $p_0$ and $p_1$ as a function of \lge, in the form

\begin{equation}
  \begin{split}
    p_0(E) &= \alpha_0 \cdot \log_{10}(E/{\rm eV}) + \beta_0, \\
    p_1(E) &= \alpha_1 \cdot \log_{10}(E/{\rm eV}) + \beta_1.
  \end{split}
  \label{eq:toy:ps}
\end{equation}

\noindent
The dotted lines in~\cref{fig:toy:param:eposlhc} show the function from~\cref{eq:toy:ps} fitted to the point obtained from CORSIKA simulations. The values of $\alpha_0$, $\beta_0$, $\alpha_1$ and $\beta_1$ for all primaries and hadronic interaction models are presented in~\cref{tab:toy}.

\begin{figure*}[]
  \subfloat[Proton]{
    \includegraphics[width=0.47\textwidth]{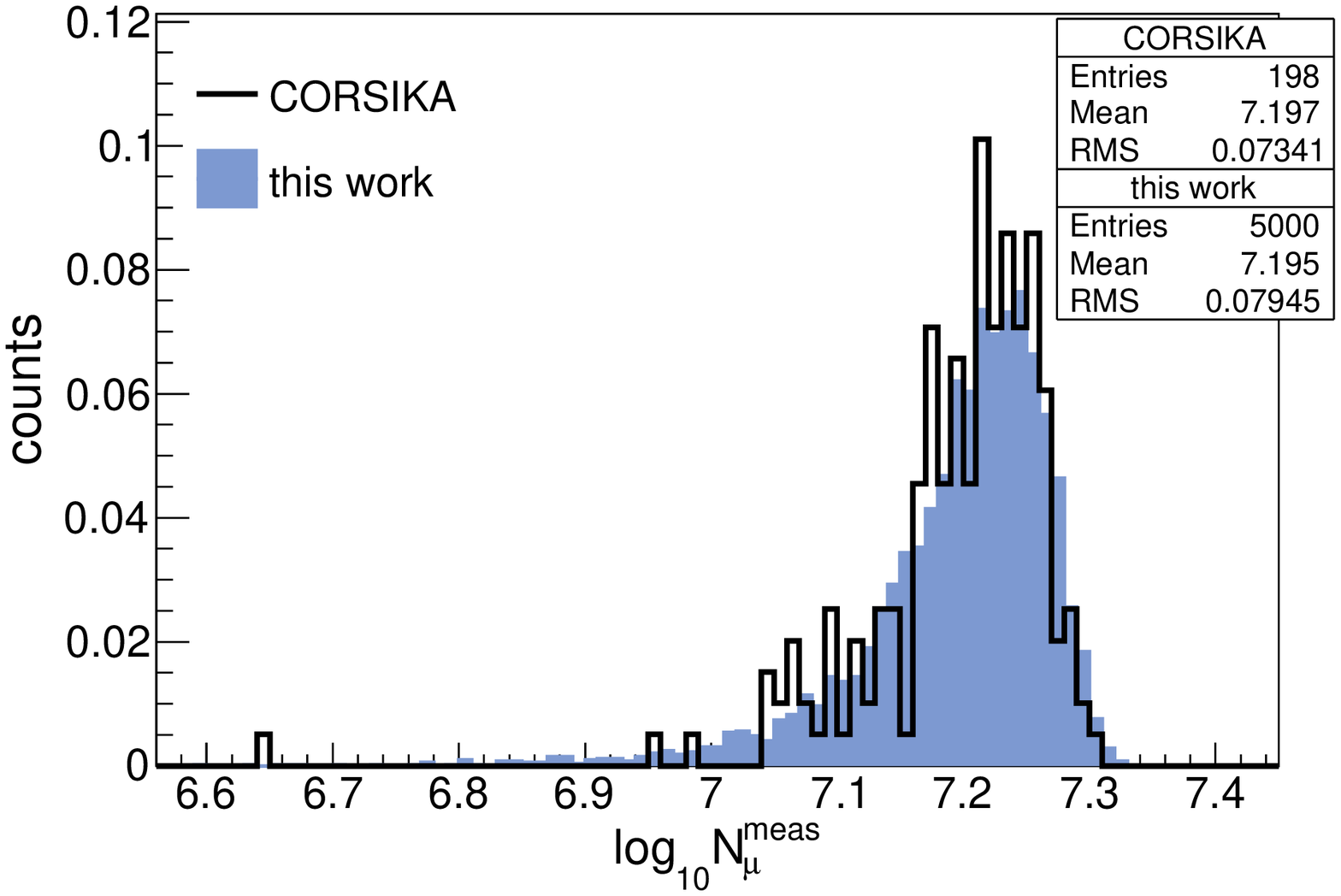}
  }
  \subfloat[Helium]{
    \includegraphics[width=0.47\textwidth]{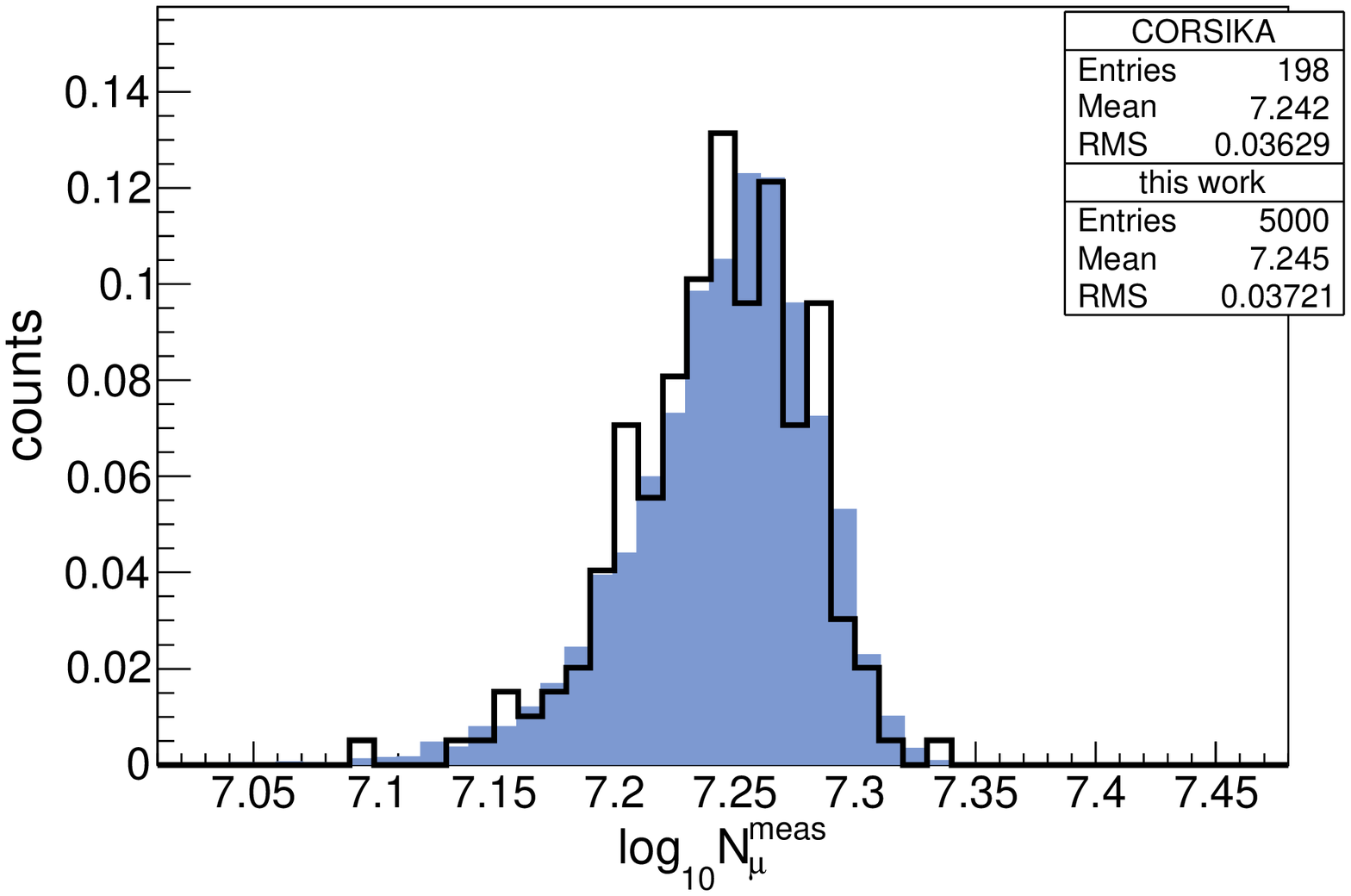}
  }

  \subfloat[Nitrogen]{
    \includegraphics[width=0.47\textwidth]{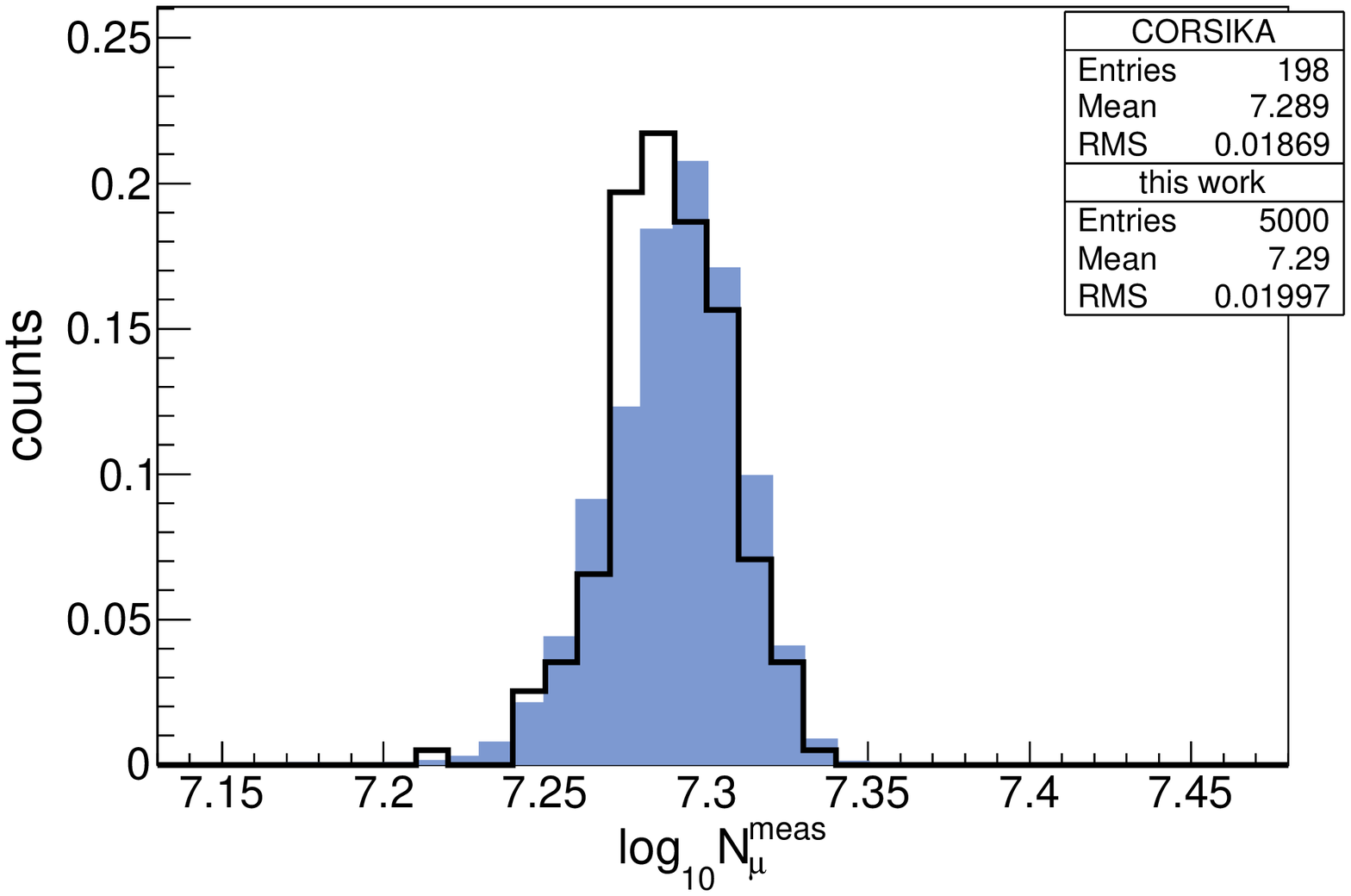}
  }
  \subfloat[Iron]{
    \includegraphics[width=0.47\textwidth]{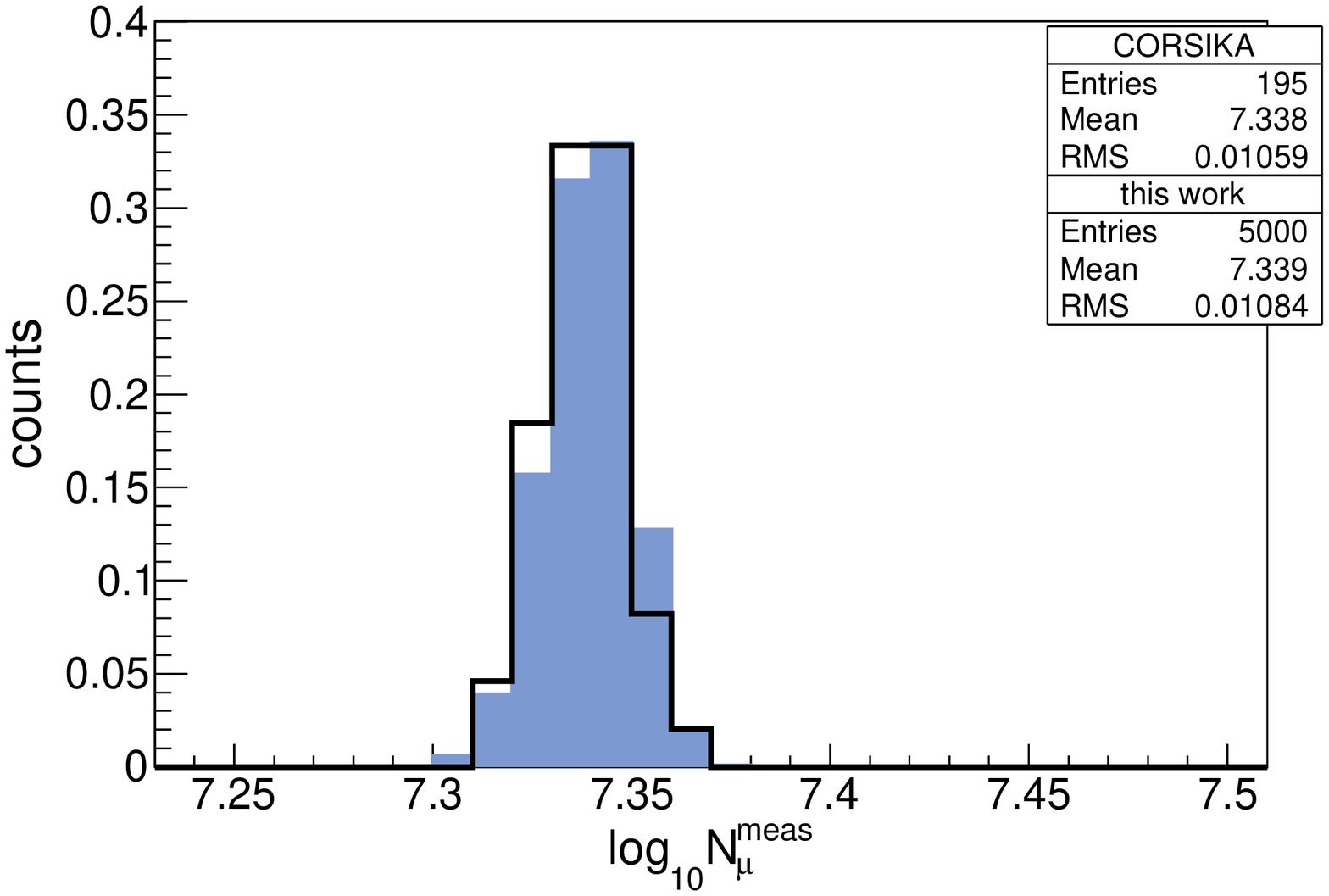}
  }

  \caption{Comparison between the normalized \lgnmuexp distributions generated by CORSIKA and by this work's algorithm. The energy is $E=10^{19.0}$ eV and the hadronic interaction model is EPOS-LHC. The primary particles are (a) proton, (b) helium, (c) nitrogen and (d) iron.}
  \label{fig:toy:dist:eposlhc}
\end{figure*}

A comparison between \lgnmuexp distributions achieved directly from CORSIKA showers and from the simulated method described in this work can be seen in~\cref{fig:toy:dist:eposlhc}. The energy is fixed at $E = 10^{19.0}$ eV and the hadronic interaction model is EPOS-LHC. The discrepancies in the mean values and in the $\sigma$ of the distributions are less than 2\% and 5\% respectively, for any combination of primary and hadronic interaction model. Indeed, considering the differences between the approaches assumed by both software, CORSIKA and CONEX, these observed discrepancies are really satisfactory. Among the several physical effects which are treated differently we can highlight the lack of geomagnetic field and muon multiple scattering in CONEX.

Although there is a small discrepancy between our method's and CORSIKA's \lgnmuexp distributions, we do not expect to find any loss in the development of this paper. This can be assured because the method proposed here is not dependent on the comparison between our simulations and any other set of simulated showers.

\begin{table*}
  \begin{center}
    \begin{tabular}{|c|c|c|c|c|c|}\hline
      had. int. model & primary & $\alpha_0$ & $\beta_0$ & $\alpha_1$(10$^{-5}$g$^{-1}$cm$^2$) & $\beta_1$(10$^{-5}$g$^{-1}$cm$^2$) \\ \hline \hline
      \multirow{4}{*}{EPOS-LHC} & proton & 0.0303 & -0.00635 & -3.20 & 38.0\\
      & helium & 0.0385 & -0.147 & -4.14 & 53.6\\
      & nitrogen & 0.0337 & -0.0721 & -3.90 & 50.5\\
      & iron & 0.0259 & 0.0613 & -3.04 & 35.8\\ \hline
      \multirow{4}{*}{QGSJetII-04} & proton & 0.0216 & 0.130 & -2.04 & 17.3 \\
      & helium & 0.0143 & 0.264 & -1.20 & 1.35 \\
      & nitrogen & 0.0167 & 0.209 & -1.60 & 9.94 \\
      & iron & 0.0245 & 0.0427 & -2.99 & 38.5 \\ \hline
    \end{tabular}
    \caption{Fitted parameters $\alpha_0$, $\beta_0$, $\alpha_1$ and $\beta_1$ given in \cref{eq:toy:ps}.}
    \label{tab:toy}
  \end{center}
\end{table*}

%\section*{References}
\bibliographystyle{elsarticle-num}
\bibliography{mylib.bib}

\newpage

%=====================================================================

%\end{linenumbers}

\end{document}